\documentstyle[12pt,epsfig,cite]{article}

\def\s#1{\setbox0=\hbox{$#1$}%
\rlap{\ifdim\wd0>.7em\kern.22\wd0\else\kern.1\wd0\fi /}#1}

\newcommand{\beq}{\begin{equation}}
\newcommand{\eeq}{\end{equation}}
\newcommand{\bea}{\begin{eqnarray}}
\newcommand{\eea}{\end{eqnarray}}

\newcommand{\tto}{\!\to\!}

\newcommand{\gsim}{\lower.7ex\hbox{$
\;\stackrel{\textstyle>}{\sim}\;$}}
\newcommand{\lsim}{\lower.7ex\hbox{$
\;\stackrel{\textstyle<}{\sim}\;$}}

\newcommand{\bibit}[1]{\bibitem{#1}}

\newcommand{\aver}[1]{\langle #1\rangle}

\newcommand{\La}{\overline{\Lambda}}

\newcommand{\Lam}{\Lambda_{\rm QCD}}
\newcommand{\mhad}{\mu_{\rm hadr}}

\newcommand{\al}{\alpha}
\newcommand{\be}{\beta}
\newcommand{\ga}{\gamma}

\newcommand{\la}{\lambda}

\newcommand{\GeV}{\,\mbox{GeV}}

\newcommand{\matel}[3]{\langle #1|#2|#3\rangle}
\newcommand{\state}[1]{|#1\rangle}

\newcommand{\eod}{\end{document}}

\newcommand{\msp}[1]{\mbox{\hspace*{#1mm}~}}

\begin{document}
\thispagestyle{empty}
\vspace*{-22mm}

\begin{flushright}
Bicocca-FT-05-25\\
CERN-PH-TH/2005-223\\
UND-HEP-05-BIG\hspace*{.08em}06\\
LPT-ORSAY~05-73\\
FTPI-MINN-05/18 \\ 
IPPP/05/62\\
DCPT/05/124\\
hep-ph/0511158\\

\end{flushright}
\vspace*{1.3mm}

\begin{center}
{\LARGE{\bf
On the nonperturbative charm effects \vspace*{4mm}\\
in inclusive \boldmath $\,B\tto X_c\,\ell\nu\,$ decays
}}
\vspace*{19mm}

{\Large{\bf I.I.~Bigi$^{\,a}$\,  
N.~Uraltsev$^{\,a,b\,*}$ and R.~Zwicky$^{\,c,d} $}} \\
\vspace{7mm}

$^a$ {\sl Department of Physics, University of Notre Dame du Lac}
\vspace*{-.8mm}\\
{\sl Notre Dame, IN 46556, USA}\vspace*{1.5mm}\\
$^b$ {\sl INFN, Sezione di Milano, Milan, Italy} \, 
\raisebox{-4pt}{{\scriptsize \hspace*{-6pt}and}} \\
{\sl CERN, Theory Division, Geneva, Switzerland} \vspace*{1.5mm}\\
$^c$ {\sl William Fine Theoretical Physics Institute, University Minnesota}
\vspace*{-.8mm}\\
{\sl Minneapolis, MN 55455, USA}\vspace*{1.5mm}\\
$^d$ {\sl IPPP, Department of Physics, University of Durham}
\vspace*{-.8mm}\\
{\sl  Durham DH1 3LE, UK}\vspace*{.5mm}\\ 

\vspace*{10mm}

{\bf Abstract}\vspace*{-1.5mm}\\
\end{center}

\noindent
We address the nonperturbative effects associated with soft charm
quarks in inclusive $B$ decays. Such corrections are allowed by
the OPE, but have largely escaped attention so far. The related
four-quark `double heavy' expectation values of the form
$\matel{B}{\bar{b}c\,\bar{c}b}{B}$ are computed in the $1/m_c$
expansion by integrating out the charm field to one and two
loops. A significant enhancement of the two-loop coefficients is
noted. A method is suggested for evaluating the expectation
values of the higher-order $b$-quark operators required to
calculate charm expectation values, free from the overly large
ambiguities of dimensional analyses. The soft-charm effects were
found generally at the level of $0.5\%$ in $\Gamma_{\rm sl}(b\tto
c)$; our literal estimate is somewhat smaller as a result of
partial cancellations. We propose a direct way to search for such
effects in the data. Finally, we discuss the relation of the soft
charm corrections in inclusive decays with the `Intrinsic Charm'
ansatz.

\setcounter{page}{0}
\vfill

~\hspace*{-12.5mm}\hrulefill \hspace*{-1.2mm} \\
{\footnotesize
\hspace*{-5mm}$^*$
On leave of absence from
St.\,Petersburg Nuclear Physics
Institute, Gatchina, St.\,Petersburg 188300, Russia}\vspace*{-7mm}
\normalsize

\newpage

%%%%%%%%%%%
\section{Introduction}
\label{INTRO}
%%%%%%%%%%%%%

Inclusive weak or radiatively induced electroweak decay rates of heavy
flavor, in particular of beauty hadrons can be rigorously treated in
the heavy quark expansion employing a local operator expansion in
powers of $1/m_Q$ (OPE). For fully integrated rates there are no
nonperturbative corrections scaling like $1/m_Q$ \cite{buv,bs}, and the
leading contributions are given to order $1/m_b^2$ in terms of the
expectation values of the kinetic and chromomagnetic heavy quark
operators.  They incorporate effects due to bound-state
dynamics in the initial state as well as the propagation of the 
{\sf hard} final state quarks through the nonperturbative hadronic medium
during the decay. In this respect they are analogous to the leading
power correction in the vacuum current correlator due to the gluon
condensate $\propto \frac{\aver{G^2}}{Q^4}$ \cite{banda}.

At order $1/m_b^3$ four-quark operators of the form
$\matel{B}{\bar{b}\Gamma q\, \bar{q}\Gamma b}{B}$ appear in the
expansion of the decay rates. They describe, for instance, weak
annihilation (or weak scattering in baryons) and Pauli interference which
distinguish the decays of hadrons with different flavors of the light
spectator quarks -- yet they generate flavor-singlet effects as
well. 
Their meaning is transparent: the propagation of soft light 
quarks in the decay is not perturbative, neither
literally nor in the duality sense, and the overall contribution from 
these kinematics should be regarded as an additional nonperturbative
input \cite{sv,WA}. It parallels the quark condensate effect in the
vacuum correlator \cite{banda}. 

Inclusive semileptonic $B$ decays offer the cleanest theoretical
environment for precision application of the OPE. They allow to
extract both the heavy quark parameters $m_b$, $m_c$, $\mu_\pi^2$
etc. and $|V_{cb}|$ or $|V_{ub}|$ with (almost)
unmatched~\footnote{The $b$ quark mass has been extracted from the
threshold cross section of $e^+e^- \tto \bar{b}b$ with similar
accuracy \cite{mbthresh}; that value is in a very good agreement with the
recent results from inclusive $B$ decays.} accuracy (see, for
instance, Ref.~\cite{buch}).
It is then essential to analyze all contributions, even small 
ones \cite{imprec}. 

The four-fermion operator $\bar{b}\gamma_k(1\!-\!\gamma_5)u
\,\bar{u}\gamma^k(1\!-\!\gamma_5)b$ explicitly appears at tree
level in the semileptonic $b\tto u$ decays \cite{WA}.  In the dominant
semileptonic mode $b\tto c$ analogous contributions from
$\bar{b}\gamma_k(1\!-\!\gamma_5)c \,\bar{c}\gamma^k(1\!-\!\gamma_5)b$
are usually neglected under the tacit assumption that $c$ quarks are
sufficiently heavy as not to induce nonperturbative
effects.\footnote{A partial exception is the $c$-quark--induced
Darwin term which, by the equations of motions, reduces to the
flavor-singlet $4\pi\alpha_s \bar{b}\gamma_0 t^a b\,\bar{q}\gamma_0
t^a q$ operator, manifesting the perturbative mixing of
$\bar{b}\gamma_k(1\!-\!\gamma_5)c \,\bar{c}\gamma^k(1\!-\!\gamma_5)b$ 
with the latter.} 
This assumption is valid only up to a point: nonperturbative charm
effects could a priori become relevant at the one percent level.

There is a complementary reason to introduce explicitly the four-quark
expectation values with the charm quark fields, which becomes manifest
once the heavy quark expansion in $b\tto c\,\ell\nu$ is extended
beyond order $1/m_b^3$. For instance, in calculating $\Gamma_{\rm
tot}$ we expand the charm quark Green function in the external
field. Similar to differentiating over $m_c^2$, this produces
increasingly infrared-sensitive integrals converging at $k_c\lsim m_c$
rather than at the hard scale of the energy release. Consequently, the
resulting expansion in general includes terms scaling like
\beq
\Gamma_0\, \mhad^{k+l} \, \frac{1}{m_b^k}\, \frac{1}{m_c^l}
\label{10}
\eeq
with $l\!>\!0$ for $k\!\ge\!3$. Just for the above reasons these terms
describe precisely the lump effect of the expectation values like
$\matel{B}{\bar{b}\Gamma c\, \bar{c}\Gamma b}{B}$. Including the
latter explicitly allows to preserve the principal advantage of the
OPE for the inclusive rates, viz.\ expansion in $1/m_b$ (or, more
generally, in the inverse energy release) free from corrections
suppressed only by the inverse charm mass.

Therefore, with the charm quark significantly lighter than the $b$
quark and only marginally heavy on the QCD mass scale, it is
advantageous to explicitly introduce the four-quark
operators with charm into the OPE and analyze their impact.
\vspace*{1.5mm}

The remainder of the paper will be organized as follows. In Sect.~2
the framework is set up which is then used to compute the charm
expectation values in the $1/m_c$ expansion to one- and two-loop
level. The expectation values of the $b$-quark operators of dimension
$7$ and $8$ are estimated, and the contributions to $\Gamma_{\rm
sl}(b\tto c)$ from both $1/m_c^2$ and $\alpha_s/m_c$ terms are
evaluated. In Sect.~3 we discuss the phenomenology of the soft-charm
effects and point out that the $q^2$-moments of the decay
distributions are rather sensitive to them. In Sect.~4 we summarize
our conclusions and dwell on a possible connection of these effects
with `Intrinsic Charm' in $B$ particles. Appendices guide through
technical details of our analysis: the conventions are described in
Appendix A, the master technique for calculation is briefly reviewed
in Appendix B and in Appendix C we describe the main steps in the
technically challenging two-loop calculations.

%%%%%%%%%%
\section{The framework}
%%%%%%%%%

The total semileptonic $b\tto c$ decay rate in the OPE is given by 
\small
\bea
\nonumber
\Gamma_{\rm sl}(b\!\to\!c)\msp{-3} &=& \msp{-3}\frac{G_{F\,}^2
m_b^5(\mu)}{192\,\pi^3}\; \raisebox{-.5mm}{\mbox{{\large$|V_{cb}|^2$}}} 
\,\left(1\!+\!A_{\rm ew}\right)\left[z_0(r)\,
(1\!+\!A_3^{\rm pert}(r;\mu))\!\left(
1\!-\!\frac{\mu_\pi^2(\mu)\!-\!\mu_G^2(\mu) \!+\! 
\frac{\rho_D^3(\mu)\!+\!\rho_{LS}^3(\mu)}{m_b(\mu)}}
{2m_b^2(\mu)}\right)\right.\\
&-& \msp{-3} \left. 
(1+A_5^{\rm pert}(r;\mu))\, 2(1\!-\!r)^4 
\frac{\mu_G^2(\mu)\!-\!\frac{\rho_D^3(\mu)\!+
\!\rho_{LS}^3(\mu)}{m_b(\mu)}}{m_b^2(\mu)}+
(1+A_D^{\rm pert}) d(r)\,\frac{ \rho_D^3(\mu)}{m_b^3(\mu)}
\right.\nonumber\\
&+& \msp{-3} \left.
32\pi^2\,(1+A_{6c}^{\rm pert}(r))(1\!-\!\sqrt{r})^2
\frac{H_c}{m_b^3(\mu)} + 
32\pi^2\,\tilde A_{6c}^{\rm pert}(r) 
(1\!-\!\sqrt{r})^2\frac{\tilde H_c}{m_b^3(\mu)}
\right.\nonumber\\
&+& \msp{-3} \left.
32\pi^2 A_{6q}^{\rm pert}(r)\frac{F_q}{m_b^3(\mu)} +
{\cal O}\left(\frac{1}{m_b^4}\right)
\right] .
\qquad \qquad 
\label{14}
\eea
\normalsize
The coefficients $A_i^{\rm pert}$ stand for perturbative $\alpha_s$
corrections and $r=(m_c/m_b)^2$, 
\bea
\nonumber
z_0(r)\msp{-4}&=&\msp{-4} 1-8r+8r^3-r^4-12r^2\ln{r}\;,\\
d(r)  \msp{-4}&=&\msp{-4}  8\ln{r} + \frac{34}{3}-\frac{32}{3}r -8 r^2
+\frac{32}{3}r^3 -\frac{10}{3} r^4\;.
\label{8}
\eea
A detailed discussion can be found
in Ref.~\cite{imprec}. We have adopted the Wilsonian prescription of
introducing an auxiliary scale $\mu$ to separate  large- and
short-distance dynamics consistently.

The last term in Eq.~(\ref{14}) proportional to $F_q$ denotes the
effect of generic $SU(3)$-singlet four-quark operators (other than
Darwin operator) of the form $\bar{b}\Gamma b\:\bar{q}\Gamma q$ with
the sum over $q=u,d,s$, and $\Gamma$ including both color and Lorentz
matrices (to the leading order in $1/m_b$ only $\gamma_0\times
\gamma^0$ or $\gamma_i\gamma_5\times\gamma^i\gamma_5$ structures
survive, but one does not need to rely on this). Their Wilson
coefficients are ${\cal O}(\alpha_s)$, and we neglect these
contributions.

Since the $b$ quark is sufficiently heavy, we consider terms only
through order $1/m_b^3$. This includes the four-quark operators with
the charm fields, yet only those without derivatives -- the latter
appear to order $1/m_b^4$ or higher. Without perturbative loop
corrections only
\beq
H_c=\frac{1}{2M_B}
\matel{B}{\bar{b}\gamma_\nu(1\!-\!\gamma_5)c\, 
\bar{c}\gamma^\rho(1\!-\!\gamma_5)b}{B}_\mu \: 
\left(-\delta^\nu_\rho\!+\!v^\nu v_\rho\right)\,, \qquad 
v_\nu = \frac{P^B_\nu}{M_B}
\label{24}
\eeq
will contribute.  This form literally implies the demarcation point
$\mu$ to be above the $c$ quark mass, in which case charm is fully
dynamical. The four-quark expectation values then are dominated by the
large perturbative piece $\propto\mu^3/6\pi^2$. It is natural to
eliminate it by evolving the theory down to smaller $\mu$ and
eventually integrating out the charm field. This procedure adds
nonperturbative corrections to be analyzed below.

Limiting ourselves to the third order in $1/m_b$ -- the leading one
for these operators -- amounts to treating them in the static limit
$m_b\tto \infty$. There are then in general four $O(3)$-invariant
parity-conserving operators possible of the form $\bar{b}b\,
\bar{c}\gamma_0c$ and $\bar{b} \vec\sigma
b\,\bar{c}\vec\gamma\gamma_5c$, with two color contraction schemes
each. In the actual semileptonic decays a certain combination enters
which can be decomposed into color-singlet and color-octet operators
in the $s$-channel:
\bea
\nonumber
H_c=\aver{O_1^c} &\msp{-3}=\msp{-3}& 
-\frac{1}{2M_B}\matel{B}{\bar{b} \gamma_i(1-\gamma_5) c \,
\bar{c}\gamma^i(1-\gamma_5)b}{B}\\
F_c=\aver{O_2^c} &\msp{-3}=\msp{-3}& 
-\frac{1}{2M_B}\matel{B}{\bar{b} \mbox{$\frac{\lambda^a}{2}$}
\gamma_i(1-\gamma_5) c \,
\bar{c}\mbox{$\frac{\lambda^a}{2}$}\gamma^i(1-\gamma_5)b}{B}\;. \qquad
\label{40}
\eea
At a high normalization point only $H_c$ enters at tree level
according to Eq.~(\ref{14}), however strictly speaking these two mix
under renormalization. For our analysis we consider, in general, the
four operators
\bea
\nonumber
O_V^s=\bar{b}b \, \bar{c}\gamma_0c  \msp{7}& & 
O_A^s=\bar{b}\vec\sigma b \, \bar{c}\vec\gamma\gamma_5 c \\
O_V^o=\bar{b}\mbox{$\frac{\lambda^a}{2}$} b \,
\bar{c}\mbox{$\frac{\lambda^a}{2}$} \gamma_0 c & &
O_A^o=\bar{b}\mbox{$\frac{\lambda^a}{2}$} \vec\sigma b \, 
\bar{c}\mbox{$\frac{\lambda^a}{2}$} \vec\gamma\gamma_5 c ,  \qquad 
\label{42}
\eea
which emerge after Fierz reordering of the operators in
Eq.~(\ref{40}):
\bea
\nonumber
O_1^c &\msp{-3}=\msp{-3}&-\frac{3}{2N_c} O_V^s + \frac{1}{2N_c} O_A^s
-3 O_V^o + O_A^o\\
O_2^c &\msp{-3}=\msp{-3}& -\frac{3}{4}(1-\mbox{$\frac{1}{N_c^2}$})
O_V^s 
+ \frac{1}{4}(1-\mbox{$\frac{1}{N_c^2}$}) O_A^s + 
\frac{3}{2N_c} O_V^o - \frac{1}{2N_c}O_A^o
\;.
\label{44}
\eea
We therefore have vector and axial currents of $c$ quarks created at
the origin (the position of the $b$ quark); both can be color
singlet or octet. 

For sufficiently heavy charm quarks we can integrate them out from the
theory and have an expansion of $H_c$ and $F_c$ (or, in general, for all the four 
operators in Eq. (\ref{44})) in terms of the local $b$ quark
operators $o_k$ containing only gluon and {\em light} quark fields:
\beq
\aver{O^c_j}= \sum_k C_{jk}(\mu)\frac{1}{2M_B}
\frac{\matel{B}{\bar{b}o_k b(0)}{B}_{\mu}}{m_c^{d_k-3}}
\;,
\label{48}
\eeq
where $d_k\!\ge\! 3$ denote the dimension of the operators $o_k$. This series is an 
expansion in inverse powers of charm quark mass.

%%%%%%%%%%%%
\subsection{Integrating out the charm quarks}
%%%%%%%%%%%%

To integrate out charm quarks from the 
operators in Eq.~(\ref{42}) we calculate the one-loop as well as the
two-loop (one hard gluon exchange) contributions.  The latter are
of the leading power in $1/m_c$, however contain an extra power of the
running coupling $\alpha_s$.

%%%%%%%%%%%%
\subsubsection{One-loop expansion}
%%%%%%%%%%

The operators in Eqs.~(\ref{42}) can generically be written as
\begin{equation}
\label{49}
J_{12} \equiv \bar b \Gamma_{\!1} b \:\bar c \Gamma_{\!2} c \;;
\end{equation}
to one loop order the dependence on $\Gamma_{\!1,2}$ totally factorizes when
the charm field is integrated out. 
Without hard gluon corrections the expansion (\ref{48})
reduces to the charm quark current in an external gluon field. 
In the notation of 
Eq.~(\ref{49})
\begin{equation}
\label{50}
J_{12} = \bar b\, \Gamma_1 \aver{\bar{c} \Gamma_2 c}_A  b\;;
\end{equation} 
where the average denotes the gluonic operator obtained from the charm 
Green function in the given external gauge field $A$.  
The required
expectation values are then given by the averages of these composite
operators over the $B$ meson states. 

In practice we use the well-elaborated Fock-Schwinger gauge method, 
which yields the result directly in terms of
gauge-covariant objects. Here we shall only state the
results.~\footnote{While the present study was being written  a paper
\cite{trine} appeared which also considered, in a different
context, the generic charm quark loop
in the gluon background to order $1/m_c^2$. We are grateful to
S.~Trine for bringing it to our attention. Comparing with the
expressions there helped us to find an inconsistency in the routine
used for transforming the results to various forms which lead to 
incorrect final expressions for the $1/m_c^2$ terms presented in the
preliminary version of this paper.} 
The method is briefly described in Appendix \ref{app:1loop} and discussed
in more detail in \cite{roman}.

The absence of the $b$ quark 
when calculating $\aver{\bar{c} \Gamma_{\!2} c}_A$ 
brings in additional symmetries and simplifies the
possible structure of the operators in the expansion. Lorentz and
gauge invariance ensure that the lowest possible operators are of the
general form $DGG$, with $D$ the covariant derivatives and $G$
the gluon field strength. Therefore
these operators are of order $1/m_c^2$. They agree with the
results presented in Ref.~\cite{trine}.
The vector color-singlet current emerges only at the 
higher order $1/m_c^4$.

In what follows we will mostly use the matrix representation; 
the covariant derivative $D_\mu \!=\! \partial_\mu \!-\!iA_\mu$ then
acts as a commutator. This sometimes will be assumed and not written
explicitly.  More definitions can be found in Appendix \ref{app:conv}.

%%%%%%%%%
\subsubsection*{Axial current}
%%%%%%%%%%%

For the axial current with arbitrary color we get to order $1/m_c^2$: 
\beq
({\cal A}_\mu)^{ba} \equiv
\aver{\bar{c}^a \gamma_\mu\gamma_5 c^b}_A=  
\frac{1}{48\pi^2 m_c^2}\,\left(
2\{
[D_\alpha, G^{\alpha\beta}],\tilde G_{\mu\beta}\} +
\{[D_\alpha, \tilde G_{\mu\beta}],G^{\alpha\beta}\}\right)^{ba}
\:,
\label{54}
\eeq
where $a$ and $b$ are fundamental (spinor) indices in the $SU(3)$ color
space. Note that $[D_\alpha, G^{\alpha\beta}]$ in the first term
reduces to the color quark current, $g_s^2 \sum_q \bar{q}\gamma_ \beta
q$ in the adjoint representation.
For a spatial Lorentz index $i$ we can write
$$
({\cal A}_i)_{ba} = \frac{1}{48\pi^2 m_c^2}\,
\left( 2\{[D_i,B_k],E_k\} + \{[D_i,E_k],B_k\}-  \{[D_k,B_i],E_k\}
+ \{[D_k,E_k], B_i\} - \right.
$$
\beq
\left.
\msp{80}\epsilon_{ikl}[D_0,E_k],E_l\} \right)_{ba}
\label{60}\;.
\eeq 

The application to the semileptonic decays does not require singling
out the color-singlet piece of the current, and the obtained general 
expression appears
most convenient. The case of the color 
singlet axial current corresponds to taking the trace over the color
indices; it results in a much simpler equation:
\beq
A_\mu^s \equiv
{\rm Tr}[{\cal A}] \equiv \aver{\bar{c}^a\gamma_\mu\gamma_5 c^a}_A= 
\frac{1}{24\pi^2m_c^2}\,
\partial_\alpha \,{\rm Tr}\,G^{\alpha\nu}\tilde G_{\mu\nu}
\label{52}
\eeq
The axial singlet current has been calculated in Ref.~\cite{mpolyak}.
Assuming that $\mu\!=\!i$ is a spatial index, the singlet current can
be  rewritten in the form 
\beq
\partial_{\alpha\,}{\rm Tr}\, G^{\alpha\nu}\tilde G_{i\nu} = 
2\,\partial_{i\,} {\rm Tr}\,( \vec E  \vec B) - 
\partial_{k\,} ({\rm Tr}\,E_k B_i - {\rm Tr}\,E_i B_k)  \: .
\label{52a}
\eeq
The expression for the general axial current can be cross-checked
\cite{roman} via the non-Abelian axial anomaly using the expansion
of the pseudoscalar density.

%%%%%%%%%
\subsubsection*{Vector current}
%%%%%%%%

For the most general color vector current we get 
$$
({\cal V}_\mu)_{ba} \equiv
\aver{\bar{c}^a\gamma_\mu  c^b}_A=  
\frac{i}{240\pi^2 m_c^2}\,\left(
13 [D_\beta, [G_{\alpha\mu}, G^{\alpha\beta}]] + 8i( [D^\alpha, 
[D^\beta,[ D_\beta, G_{\alpha\mu}]]]\right.
$$
\beq
\left.
\msp{40}-4i[D^\beta, [D^\alpha, [D_\beta, G_{\alpha\mu}]]] \right)_{ba}
\;.
\label{62}
\eeq
It is readily verified that this current is covariantly conserved,
$D^\mu {\cal V}_\mu \!=\! 0$ \cite{roman} as required by gauge
invariance.  Note that we have dropped the leading term $D^\nu
G_{\nu\mu}\!=\!g_s^2 \sum_q \bar{q} \gamma_\mu q$ with a coefficient
$\propto \!\ln {\frac{\Lambda^2_{\rm UV}}{m_c^2}}$, since it is
accounted for in the Darwin operator in the width at order
$\frac{1}{m_b^3}\frac{1}{m_c^0}$. The pure octet component of the 
current can be obtained by varying the gauge field effective action
with respect to $A^d_\mu$ which yields ${\rm Tr}\,({\cal V}_\mu t^d)$;
this is explicitly verified in \cite{roman}. By setting $\mu \!=\! 0$
-- the only component surviving the static limit of the $b$ quarks --
we can write the expression for the current as
\begin{eqnarray}
({\cal V}_0)_{ba} 
=13 \left( D\!\cdot\!(E\!\times\!B)+  
D\!\cdot\!(B\!\times\!E)\right) + 
4i(2D_\nu D^\beta -D_\nu D^\beta)  D_\beta E_k)  
\label{66}
\end{eqnarray}
(here the usual commutators are assumed when we
symbolically write products of covariant derivatives).

The vector color-singlet current vanishes at order $1/m_c^2$ 
as follows from an explicit calculation. This was anticipated in 
Ref.~\cite{mpolyak} based on $C$-parity arguments. The leading 
term appears at order $1/m_c^4$ and has the general form $DGGG$.
This current describes the effective local interaction
of the photon with the gluon fields  (in the
vacuum) due to heavy quarks, and is of independent
interest. It can be written as a total derivative:  
\beq
{\cal V}^s_\mu \equiv
\aver{\bar{c}^a\gamma_\mu  c^a}_A=-  
\frac{1}{1440\pi^2 m_c^4}\, \partial^\alpha
\left(
14 \{G_{\alpha\beta} G^{\beta\gamma} G_{\gamma\mu} \}_d 
+ 5\{G_{\gamma\beta} G^{\beta\gamma} G_{\alpha\mu} \}_d \right)
\;,
\label{70}
\eeq
where $\{\dots\}_d$ means that the symmetric matrix product
$\sim d_{abc}$ of the three matrices must be taken. This expression
vanishes for the $SU(2)$ gauge group, as expected 
from $G$-parity \cite{mpolyak,roman}.
Gauge invariance ensures this current to be conserved,  
$\partial^\mu {\cal V}^s_\mu\!=\!0$,
which is evident from the symmetry properties of the field 
strength tensors.
The two terms, however, do not vanish for the Abelian $U(1)$ 
group; they can be obtained from the Euler-Heisenberg 
effective action as a variation with respect to the gauge 
field \cite{roman}. Eq.~(\ref{70}) therefore represents 
the minimal non-Abelian extension
of the Abelian current. Once again anticipating that 
$\mu \!=\! 0$ in the expectation values,  
the two terms above can be written as 
\begin{eqnarray}
{\cal V}^s_0 = \partial^\alpha (
14 \{G_{\alpha\beta} G^{\beta\gamma} G_{\gamma 0} \}_d 
&+& 5\{G_{\gamma\beta} G^{\beta\gamma} G_{\alpha 0} \}_d ) = \nonumber  \\[2mm]
& & \partial_a \{ 4(E_a E^2) -24(E_aB^2)-7(B_a(E\cdot B) \}_d \,.
\end{eqnarray}
The one-loop contributions considered above add up to 
\beq
H_c^{\rm 1-loop} = -\frac{3}{2} \aver{\bar{b}{\cal V}_0 b} + 
\frac{1}{2}\aver{\bar{b}\vec \sigma 
\vec {\cal A} b}\,.
\label{262}
\eeq

%%%%%%%%%%%
\subsection{Two-loop expansion to order \boldmath$\alpha_s(m_c)/m_c$}
%%%%%%%%%%

Accounting for color Coulomb effects of the $b$ quark
reduces the symmetries of the $c$ quark interaction with the soft
gluons and allows more operators to appear. In particular, gluon
operators of dimension $4$ suppressed only as $1/m_c$ can appear for
both vector and axial currents at the price of an extra loop and
the associated factor $\alpha_s(m_c)$. There are six such effective 
operators allowed on general grounds by $P$ and $T$ 
parities:\footnote{The operators with the chromomagnetic field were
left out in Ref.~\cite{imprec}.}
\beq
\addtolength{\arraycolsep}{6pt}
\renewcommand{\arraystretch}{1.5}
\begin{array}{ll}
E_{\delta} =  \frac{1}{2N}\delta_{ab} \bar{b} E^a\!\cdot\! E^b b  
& \quad B_{\delta} =  \frac{1}{2N}\delta_{ab} \bar{b} B^a\!\cdot\! B^b b \\
E_{d} =  \frac{1}{2}d_{abc} \bar{b} E^a\!\cdot\! E^b  t_c b  
& \quad B_{d} =  \frac{1}{2}d_{abc} \bar{b} B^a\!\cdot\! B^b t_c b \\
E_{f} = -\frac{1}{2}f_{abc} \bar{b} E^a\!\times\! E^b \!\cdot\! \sigma  t_c b  
& \quad B_{f} = -
\frac{1}{2}f_{abc} \bar{b} B^a\!\times\! B^b  \!\cdot\! \sigma t_c b \;.
\end{array}
\addtolength{\arraycolsep}{-6pt}
\renewcommand{\arraystretch}{1.}
\label{80}
\end{equation}
The first four operators appear in the vector current while the last
two are generated by the axial current; the former lead to the
spin-singlet operator and the latter to the operator proportional to
the $b$ quark spin.

Conventional wisdom would tell us that the extra loop and the
associated suppression by the perturbative factor $\sim\! 
\alpha_s/\pi$ would yield a very small factor. It was conjectured in
Ref.~\cite{imprec} that the enhancement factor $\frac{m_c}{\mhad}$ in
the matrix element would fall short of compensating for
it. Nevertheless, we have calculated these contributions and actually
found a significant enhancement of the perturbative corrections over
the naive loop counting rules. It is apparently associated with the
physics of the static Coulomb field: the typical coefficient contains
$\pi^2$ on top of the usual loop factor $\alpha_s/\pi$. Therefore
these higher order contributions may actually dominate unless a
cancellation among different operators at this order takes place.

Explicit calculations yield to order $\alpha_s/m_c$: 
\bea
\nonumber
\aver{O^s_V} &=& \frac{\alpha_s}{2^73^2\, \pi\,m_c}(14 \aver{B_d}+
23 \aver{E_d}) \\
\nonumber
\aver{O^o_V} &=& \frac{\alpha_s}{2^{11} 3^3 \,\pi\,m_c}
(4708 \aver{B_d}- 1511 \aver{E_d} + 9752 \aver{B_\delta} -2470 
\aver{E_\delta}) \\
\nonumber
\aver{O^s_A} &=& \frac{\alpha_s}{2^63^25 \,\pi\,m_c}(-181 \aver{B_f}-
185 \aver{E_f}) \\
\aver{O^o_A} &=& \frac{\alpha_s}{2^93^3 \,\pi\,m_c}(-422 
\aver{B_f}-295 \aver{E_f})
\label{90}
\eea
The average denotes $\aver{X} \equiv \frac{1}{2m_B}\matel{B}{X}{B}$.
Using Eqs.~(\ref{44}) we obtain 
\bea
\nonumber
H_c^{\rm 2-loop} \msp{-4}&=&\msp{-4} \frac{\alpha_s}{2^{12}3^3 5\,
\pi m_c}[\,74100 \aver{E_\delta}  + 
39810 \aver{E_d} - 17720 \aver{E_f}  - 292560 \aver{B_\delta}
 - 144600 \aver{B_d} \\
\nonumber
& & \msp{36.5} -  22672 \aver{B_f}\,] \\
\nonumber
\msp{-4}&=&\msp{-4}
\frac{\alpha_s}{2^{12}3^4 5\,\pi m_c}[\,-121928 \aver{2(B_\delta+B_d)-B_f}
-189944 \aver{B_\delta+B_d+B_f} \\
\nonumber
&&\msp{17} + 57530 \aver{2(E_\delta+E_d)-E_f}+
4370 \aver{E_\delta+E_d+E_f} \\[0.15cm]
&&\msp{17} -273375 \aver{B_\delta-E_\delta}- 
170505\aver{B_\delta+E_\delta}\,] \;.
\label{92}
\eea
The representation above is adapted to the estimates of the matrix
elements given in the next section.

%%%%%%%%%%%%%%%
\subsection{Estimates of the expectation values}
%%%%%%%%%%%%%

Our goal is to provide realistic estimates of the scale of the
expectation values of the operators obtained by integrating out the
charm field and to get a better idea of the importance of the
nonperturbative charm effects when they are treated in the $1/m_c$
expansion. Estimating the higher-dimensional expectation values is a
notoriously difficult problem, and our primary task is to elaborate a
method free from arbitrarily appearing huge factors like powers of
$4\pi$ which all too often plague meaningfulness of `Naive Dimensional
Analyses'. Since the two-loop induced $D\!=\!7$ operators have large
Wilson coefficients, we start out scrutinizing these contributions.
We have elaborated more certain estimates of five out of six $D\!=\!7$
operators in Eqs.~(\ref{80}). Our estimates for the one-loop $D\!=\!8$
and higher operators may be less accurate, yet indicate that their
effect is presumably small enough not to introduce significant
uncertainties in $\Gamma_{\rm sl}(B)$.

%%%%%%%%%%%%%
\subsubsection{Two-loop $1/m_c$ expectation values}
%%%%%%%%%%%%%%
 
The expectation values of $D\!=\!7$ operators in Eqs.~(\ref{80}) have
been denoted $\aver{E_s}$, $\aver{E_\delta}$, ..., respectively (the
$B^*$ expectation values are related to them in an obvious way, but we
do not need them). Let us first consider the `color-through' operators
which do not have the color trace for gauge fields separately, and we
start with those containing the chromomagnetic field. In what follows
we often pass to the quantum mechanical notations. For instance, we
assume
\beq
\bar{b} \pi_k \pi_l...\pi_j b \; \longrightarrow \; \pi_k
\pi_l...\pi_j\vert_{\rm QM} \msp{10} \pi_k=
-i D_k, \;\; \pi_0=i D_0\!-\!m_b\;.
\label{120}
\eeq
Indeed, as a general rule $\int d^3x \,d^3y \: \bar{b} \pi_k
b(0,\vec{x})\: \bar{b} \pi_l b(0,\vec{y}) = \int d^3y \,\bar{b} \pi_k
\pi_l b(0,\vec{y})$ holds for static quarks, since at equal time
$b^i(0,\vec{x})\bar{b}^j(0,\vec{y})=
\delta^3(\vec{x}-\vec{y})\delta^{ij}$ ($i,j$ are color indices), and
for the quantum mechanical local heavy quark operator $O$ the notation in second quantization reads as 
\beq
O_{\rm QM} = \int d^3\!x  \, \bar{b} Ob(0,\vec{x})\;. 
\label{122}
\eeq
The operator $\pi_0$ acting on the right corresponds 
to $-\cal{H}$, or more accurately, it plays the role of  
the commutator of $-\cal{H}$ with what follows it. 

We estimate  the $B$-field expectation values using a generalization
of the factorization ansatz. More precisely, it takes the form of
the saturation by the (multiplet of) the ground heavy quark
states. For a general analysis, it is convenient to consider the
static limit, where the spin of the heavy quarks gets decoupled. 
In this world 
the usual $B$ and $B^*$ mesons belong to a single spin-$\frac{1}{2}$ 
hadron state $\state{\Omega_0}$, see Ref.~\cite{rev}, with  
\beq
\matel{\Omega_0}{\bar{b} \pi_j \pi_k b}{\Omega_0}= \frac{\mu_\pi^2}{3}
\delta_{jk} \Psi_0^\dagger \Psi_0 
-\frac{\mu_G^2}{6}\Psi_0^\dagger \sigma_{jk}\Psi_0, \msp{10}
\sigma_{jk}=i\epsilon_{jkl}\,\sigma_l\;.
\label{124}
\eeq
Setting
\beq
\matel{\Omega_0}{B_j B_k }{\Omega_0}\simeq 
\sum_\lambda \matel{\Omega_0}{B_j}{\Omega_0^\lambda}
\matel{\Omega_0^\lambda}{B_k }{\Omega_0}
\label{126}
\eeq
where we have shown explicitly the polarization of the intermediate
state, we get
\beq
\matel{\Omega_0}{B_j B_k }{\Omega_0}\simeq 
\frac{(\mu_G^2)^2}{9}\Psi_0^\dagger \sigma_j\sigma_k\Psi_0\;.
\label{128}
\eeq
Translating this relation to the world of actual spinor $b$ quarks is
simple. For instance, $\vec\sigma=-2\vec j$, and in $B$ meson $\vec j
= -\vec s_Q$, where $\vec s_Q$ and $\vec j$ are the spin and  the
angular momentum of the the $b$ quark and of the light cloud,
respectively. With $\vec j^2=\vec s_Q^2=\frac{3}{4}$ and $\vec j \vec
\sigma_Q =-\frac{3}{2}$ we have 
\bea
\nonumber
\aver{B_\delta}+\aver{B_d}\msp{-4}&=&\msp{-4} 
\matel{B}{\bar{b}\vec{B}^{\,2} b}{B}\simeq
\frac{(\mu_G^2)^2}{3} \simeq 0.041\GeV^4\\
\aver{B_f}\msp{-4}&=&\msp{-4}
\matel{B}{\bar{b}\vec{B}\!\times\!\vec{B}
\cdot \vec{\sigma} b}{B}\simeq \frac{2(\mu_G^2)^2}{3}\simeq
0.082\GeV^4 .
\label{130}
\eea
This is clearly consistent with the direct factorization for the
invariant combination 
\beq
\aver{B_\delta}+\aver{B_d}+\aver{B_f}=
\matel{B}{\bar{b}(\vec{\sigma}\vec{B})^{2} b}{B}\simeq
\matel{B}{\bar{b}\vec{\sigma}\vec{B} b}{B}^2=(\mu_G^2)^2\; . 
\label{132}
\eeq
The combination
$\aver{2(B_\delta\!+\!B_d)\!-\!B_f}=\frac{1}{3}\matel{B}{\bar{b}\,
(\vec{\sigma}\!\times\!\vec{B}+2i\vec{B})(\vec{\sigma}\!\times\!\vec{B}-2i\vec{B})\, b}{B}$ vanishes in the
factorization approximation since 
$(\vec{B} \!\times\!\vec{j}\!-\!i\vec{B})
\state{\Omega_0}=0$ (in general, $\vec{j}$ does not commute 
with $\vec{B}$, yet its commutator vanishes if projected onto the ground state 
$\Omega_0$). 
Since the two linear combinations are squares of the above operators, 
the factorization estimate gives a lower 
bound for both expectation values.

A similar ground-state factorization cannot be used for the
operators with the chromoelectric field: in the saturation of their
product 
\beq
\matel{\Omega_0}{E_j E_k }{\Omega_0}= \sum_n \matel{\Omega_0}{E_j}{n}
\, \matel{n}{E_k }{\Omega_0}
\label{140}
\eeq
only the $P$-wave states $\state{n}$ with parity opposite to that of 
$\Omega_0$ contribute. However, since $E_k=i[\pi_0,\pi_k]$
we have 
\beq
\matel{\Omega_0}{E_k }{P_n}= i \varepsilon_n 
\matel{\Omega_0}{\pi_k}{P_n}, 
\msp{10} \varepsilon_n= M_n\!-\!M_{\Omega_0}.
\label{142}
\eeq
Moreover, $P_n$ can be either $j\!=\!\frac{1}{2}$ or
$j\!=\!\frac{3}{2}$ states. The $\frac{1}{2}$-states are produced by
the (pseudo)scalar product $\vec\sigma\vec{E}$; the operator 
$\vec{\sigma}\!\times\!\vec{E}\!-\!2i\vec{E}$ creates 
$\frac{3}{2}$-states. Paralleling the derivation of Ref.~\cite{rev} 
we use 
\bea
\nonumber
\matel{\phi^{(n)}}{\pi_j }{\Omega_0} \msp{-4}&=&\msp{-4}
\varepsilon_n \tau_{1/2}^{(n)} \phi^{(n)^\dagger}\sigma_j \Psi_0\\
\matel{\chi^{(m)}}{\pi_j }{\Omega_0} \msp{-4}&=&\msp{-4}
\varepsilon_m \frac{\tau_{3/2}^{(m)}}{\sqrt{3}} 
\chi_j^{(m)^\dagger}\Psi_0\;,
\label{1}
\eea
where the spinor $\phi^{(n)}$ and the Rarita-Schwinger spinor
$\chi_j^{(m)}$ (obeying $\sigma_i \chi_i^{(m)}\!=\!0$) describe
wavefunctions of the  $j\!=\!\frac{1}{2}$ and  $j\!=\!\frac{3}{2}$
excited $P$-wave states, respectively. For the chromoelectric field we
then  
simply get extra factors of $\varepsilon^2$,
$$ 
\matel{\Omega_0}{\bar{b} E_j E_k b}{\Omega_0}= 
$$
\beq \left[2
\sum_m \,\varepsilon_m^4 |\tau_{3/2}^{(m)}|^2 + 
\sum_n \,\varepsilon_n^4 |\tau_{1/2}^{(n)}|^2 \right]
\delta_{jk}\Psi_0^\dagger \Psi_0 - 
\left[
\sum_m \,\varepsilon_m^4 |\tau_{3/2}^{(m)}|^2 - 
\sum_n \,\varepsilon_n^4 |\tau_{1/2}^{(n)}|^2 \right]
\Psi_0^\dagger \sigma_{jk}\Psi_0\;,
\label{146}
\eeq
compared to the sum rules for the products in Eq.~(\ref{124}):
\beq
\frac{\mu_\pi^2}{3}= 
2\sum_m \,\varepsilon_m^2 |\tau_{3/2}^{(m)}|^2 + 
\sum_n \,\varepsilon_n^2 |\tau_{1/2}^{(n)}|^2 , \qquad 
\frac{\mu_G^2}{3}= 
2\sum_m \,\varepsilon_m^2 |\tau_{3/2}^{(m)}|^2 -
2 \sum_n \,\varepsilon_n^2 |\tau_{1/2}^{(n)}|^2 \;.
\label{148}
\eeq
The result can be anticipated knowing the above
decomposition into the $\frac{1}{2}$- and $\frac{3}{2}$-projectors
leading to a relation similar to the case of chromomagnetic field
operators: 
\beq
2(E_\delta\!+\!E_d)\!-\!E_f=\frac{1}{3}\bar{b}\,
(\vec{\sigma}\!\times\!\vec{E}\!+\!2i\vec{E})\,
(\vec{\sigma}\!\times\!\vec{E}\!-\!2i\vec{E})\,b\,, \qquad
E_\delta\!+\!E_d\!+\!E_f=\bar{b}\,(\vec\sigma\vec{E})^2\,b\;.
\label{149}
\eeq
The former operator is saturated by $\frac{3}{2}$-states and the
latter by $\frac{1}{2}$-states. The two sum rules stemming from 
Eq.~(\ref{146}) can be written as 
\bea
\aver{2(E_\delta\!+\!E_d)-E_f} \msp{-4}&=&\msp{-4} 
\bar \varepsilon_{3/2}^2 (2\mu_\pi^2+\mu_G^2)
\label{150}\\
\aver{E_\delta+E_d+E_f} \msp{-4}&=&\msp{-4} 
\bar \varepsilon_{1/2}^2 (\mu_\pi^2\!-\!\mu_G^2)
\label{152}
\eea
where $\bar \varepsilon_{1/2, 3/2}$ denote the average excitation energies
for the two families. 

In a direct analogy to the ground-state factorization ansatz for the
chromomagnetic field operators given above one retains only 
the contribution of the lowest $P$-wave states for both
$\frac{3}{2}$- and $\frac{1}{2}$-families, $m\!=\!n\!=\!1\,$ 
in the sum over the intermediates states. This amounts to
equating the mass gaps $\bar \varepsilon$ in 
Eqs.~(\ref{150}),(\ref{152}) to the lowest
corresponding excitation energy, $\bar \varepsilon_{3/2}=
\varepsilon_{3/2}^{(1)}\equiv\varepsilon_{3/2} $ and  
$\bar \varepsilon_{1/2}=
\varepsilon_{1/2}^{(1)}\equiv\varepsilon_{1/2}$, and  
yields a lower bound which is simultaneously the factorization
estimate. 

The hierarchy for the operators with the
chromoelectric field appears to be opposite to what we get 
in case of the chromomagnetic field. The experimental 
evidence that $\mu_\pi^2\!-\!\mu_G^2
\!\ll\!\mu_\pi^2$ -- i.e.,  the proximity to the so-called `BPS limit'
\cite{BPS} -- implies
that the combination $2(E_\delta\!+\!E_d\!+\!E_f)$ must be particularly
suppressed: for in the BPS limit all $\tau_{1/2}$ vanish.
The `BPS' suppression of 
$\aver{E_\delta\!+\!E_d\!+\!E_f}$ parallels the smallness of 
nonfactorizable average 
$\aver{2(B_\delta\!+\!B_d)\!-\!B_f}$, 
while the saturation of $\aver{2(E_\delta\!+\!E_d)-E_f}$ by 
the lowest $\frac{3}{2}$ $P$-wave excitation is
a counterpart of the ground-state factorization for
$\aver{B_\delta\!+\!B_d\!+\!B_f}$.

It is interesting to note that the saturation of the small velocity
(SV, Shifman-Voloshin)
sum rules by the lowest intermediate state holds with amazing
accuracy \cite{lebur} in the exactly solvable 't~Hooft model 
\cite{thooftmodel}. Without spin in two-dimensional QCD there exists 
only a single family as analogues  of the $P$-wave states. 
It is intriguing that the latest experiments seem
to indicate a similar good saturation in actual QCD (for a recent
discussion see Ref.~\cite{qcd04}), although the data so far allow to
address this question with a reasonable accuracy only for the
$\frac{3}{2}$-channel. 

The ground-state saturation for the magnetic field like used in
Eq.~(\ref{126}) cannot be studied in two-dimensional models in view of
the single space dimension. For the symmetric products of the momentum
operators, viz.\ the kinetic operators, the ground-state saturation is
less accurate, however the deviation again is almost completely
generated by the first radial excitation.

Due to the proximity to the `BPS' regime, $\mu_\pi^2\!-\!\mu_G^2
\!\ll\!\mu_\pi^2$, the estimate of the suppressed combination
$E_\delta\!+\!E_d\!+\!E_f$ has significant relative uncertainty,
yet the absolute uncertainty must be small. The smallness of its
coefficient in $H_c$, Eq.~(\ref{92}) makes the uncertainty unimportant
in practice.
For numerical estimates here and in what follows we use the following 
values of
the heavy quark parameters $\La\!\simeq\!0.63\GeV$, 
$\mu_\pi^2\!\simeq\!0.4\GeV^2$, $\mu_G^2\!\simeq\!0.35\GeV^2$,
$\rho_D^3\!\simeq\!0.2\GeV^3$, $\rho_{LS}^3\!\simeq\!-0.15\GeV^3$, while
$\varepsilon_{3/2}\!\simeq\!0.4\GeV$ and
$\varepsilon_{1/2}\!\simeq\!0.35\GeV$ yielding literally 
\bea
\nonumber
\aver{2(E_\delta\!+\!E_d)\!-\!E_f} \msp{-4}& \simeq &\msp{-4} 0.18\GeV^4\\
E_\delta+E_d+E_f\msp{-4}& \simeq &\msp{-4} 0.006\GeV^4.
\label{160}
\eea
Equating $\varepsilon_{3/2}$ and $\varepsilon_{1/2}$ results in even
simpler relations:
\beq
E_\delta+E_d \simeq \varepsilon^2 \mu_\pi^2, \qquad 
E_f \simeq -\varepsilon^2 \mu_G^2\;.
\label{162}
\eeq

The operators where the gauge fields $E^2$ or $B^2$ are in a color
singlet are less certain. This is a counterpart of the observed
pattern discussed recently in Ref.~\cite{shifvaindiq}: 
factorization often fails where vacuum scalar contributions are
possible. We nevertheless can estimate the combination
$B_\delta\!-\!E_\delta$ since it represents (in the chiral limit for
light quarks) the expectation value of the density at the origin of the
trace of the energy-momentum tensor $\theta_{\mu\mu}$ of the light
degrees of freedom, nonvanishing due to the scale anomaly: 
\beq
\theta_{\mu\mu}= \frac{\beta(\alpha_s)}{8\pi\alpha_s^2} 
\,{\rm Tr}\, G_{\mu\nu}^2.
\label{170}
\eeq
On the other hand, the space integral of this
density over the $B$ meson volume amounts to $\bar\Lambda\!=\!M_B\!-\!m_b$ \cite{optical}:
\beq
\bar\Lambda= \frac{\beta(\alpha_s)}{8\pi\alpha_s^2} 
\,\matel{B}{{\rm Tr}\, G_{\mu\nu}^2(0)}{B} .
\label{172}
\eeq
Therefore, by dimensional estimates we would have 
\beq
B_\delta\!-\!E_\delta \approx
-\frac{16\pi^2}{27}\frac{\bar\Lambda}{\frac{4}{3}\pi R_0^3}\, ,
\label{174}
\eeq
where $\frac{4}{3}\pi R_0^3$ is the characteristic volume of the $B$
meson bound state, $R_0\!\sim\!1/\mhad$. To make the estimates of the
effective volume less vulnerable against arbitrariness in the powers
of $2\pi$ inherent in translating energy scales into inverse 
characteristic distances, we use the refined approach based on the
useful relation derived in Ref.~\cite{four}. Namely, for a
color singlet light-field operator $A$ a general relation holds
\beq
\matel{B}{\bar{b} A b(0)}{B}= \int \frac{{\rm d}^3\vec{q}}{(2\pi)^3}
{\cal F}_A(\vec{q}\,)\;,
\label{176}
\eeq
where ${\cal F}_A$ denotes the transition formfactor of the operator
$A$ alone:
\beq
\matel{B(\vec{q}^{\,})}{A(0)}{B_{\vec{p}=0}} = {\cal F}_A(\vec{q}\,)\;.
\label{178}
\eeq
We apply this relation to the energy-momentum trace operator, 
$$
A=\frac{\beta(\alpha_s)}{8\pi\alpha_s^2} \,{\rm Tr}\, G_{\mu\nu}^2.
$$
Eq.~(\ref{172}) then fixes ${\cal F}_{G^2}(0)=\La$. Assuming 
${\cal F}_{G^2}(\vec{q}^{\,2})={\cal F}_{G^2}(0)
e^{-\vec{q}^{\,2}/M^2}$ with $M^2\!\simeq\!0.5\GeV^2$ and 
$\La\!\simeq\!0.63\GeV$ we get 
\begin{eqnarray}
\frac{1}{2M_B}\matel{B}{\bar{b} 
\frac{\beta(\alpha_s)}{8\pi\alpha_s^2} \,{\rm Tr}\,
G_{\mu\nu}^2 b(0)}{B}&\approx& \frac{1}{8\pi^{3/2}} (M^2)^{3/2}
\frac{1}{2M_B}\matel{B}{\frac{\beta(\alpha_s)}{8\pi\alpha_s^2} 
\,{\rm Tr}\, G_{\mu\nu}^2(0)}{B} \nonumber \\[0.2cm]
&\simeq& \frac{1}{8\pi^{3/2}} (M^2)^{3/2}\La \simeq 0.005\GeV^4,
\label{180}
\end{eqnarray}
and 
\beq
\aver{B_\delta\!-\!E_\delta}\simeq -\frac{16\pi^2}{27}
\frac{1}{2M_B}\matel{B}{\bar{b} 
\frac{\beta(\alpha_s)}{8\pi\alpha_s^2} \,{\rm Tr}\,
G_{\mu\nu}^2 b(0)}{B}\approx -\frac{2}{27}
\sqrt{\pi}(M^2)^{3/2}\La \simeq -0.03\GeV^4
\label{181}
\eeq

Let us note that the `vacuum factorization' estimate 
$$
\matel{B}{\bar{b}b(0)}{B}\,
\aver{\mbox{$\frac{\beta(\alpha_s)}{4\alpha_s}$} \,{\rm Tr}\,
G_{\mu\nu}^2}_0 = \aver{\mbox{$\frac{\beta(\alpha_s)}{4\alpha_s}$} \,{\rm Tr}\,
G_{\mu\nu}^2 }_0
$$ 
is inappropriate since it represents the disconnected piece of the
matrix element. For this reason it has, e.g., a higher scaling in the
number of colors, $N_c^1$, while the effects we discuss scale only
like $N_c^0$. We also note that our estimate is negative, necessarily
of the opposite sign to the vacuum expectation value of the gluon
condensate. We interpret this as the fact that nonperturbative
configurations lowering the vacuum energy are suppressed inside $B$
mesons leading to a higher energy density, well in agreement with
intuition and in the spirit of the assumptions of the traditional bag
models for hadrons \cite{bag}. On the other hand, our estimate
(\ref{180}) taken at face value suggests that the decrease amounts to
a significant fraction of the vacuum density, while in the framework
of the conventional $1/N_c$ it would be suppressed as long as the
number of colors is large enough. Once again, this difference is in
line with the picture developed recently and discussed in
Ref.~\cite{shifvaindiq}.

We were not able to come up with an equally justified estimate of the
$B$ meson expectation value of the complementary color-separated
contribution $\bar{b}\,{\rm Tr}(B^2\!+\!E^2)) b$ and have to rely on
general scale estimates of its possible magnitude. We choose to
parametrize $\aver{B_\delta\!+\!E_\delta}$ in terms of 
$\aver{B_\delta\!-\!E_\delta}$, 
$$
\aver{B_\delta\!+\!E_\delta}\equiv -h_+\aver{B_\delta\!-\!E_\delta}\;;
$$
Since we speak here about nonperturbative effects and the operator in
question is spin-independent, we may expect quantitatively that the
contribution of the chromoelectric field (negative in the vacuum) is
suppressed in the $B$ meson, thus increasing $(B^2\!+\!E^2)(0)$. The
qualitative arguments in the spirit of the bag models suggest that
$\aver{B_\delta\!+\!E_\delta} \!<\! |\aver{B_\delta\!-\!E_\delta}|$
since they assume the nonperturbative fields to be suppressed in
hadrons, therefore yielding $0\!<\!h_+\!<\!1$. However, there is a
significant chromomagnetic field in $B$ meson associated with the spin
of the light degrees of freedom, $\mu_G^2\!\simeq\!0.35\GeV^2$,
therefore it should be admitted that such an argument may not fully
apply to the chromomagnetic field operators. Assuming naive positivity
of the colorless operators ${\rm Tr}\, \vec{B}^{\,2}$ and ${\rm Tr}\,
\vec{E}^{\,2}$ one would be led to conclude that $h_+\!>\!1$; however,
in the case of hadrons with a heavy quark the renormalization
properties of these operators change, and the naive positivity may be
lost.

The tables below summarize our estimates of the $D\!=\!7$ expectation
values. We show the separate matrix elements for the combinations
naturally appearing in our approach as described above, and what
follows from this for the four operators in Eq.~(\ref{90}). The latter
numbers assume $\alpha_s(m_c)\!=\!0.3$ and $m_c\!=\!1.15\GeV$ (a
short-distance Euclidean mass is adequate here). In Table~2 we also
show the estimated contribution to the total $B\tto X_c\,\ell\nu$
decay rate. Taking the numbers at face value we find a strong
cancellation among these contributions to $\Gamma_{\rm sl}$,
$-0.0007(1+h_+)$; we consider such a degree of suppression accidental
and not representative.

\begin{table}[!h]
\begin{center}
\begin{tabular}{|l|l|l|l|}
\hline
\hfill $32\pi^2\aver{O_V^s}$ \hfill  & \hfill 
$32\pi^2\aver{O_V^o}$ \hfill & 
\hfill $32\pi^2\aver{O_A^s}$ \hfill &  
\hfill $32\pi^2\aver{O_A^o}$  \hfill \\
\hline
$0.043-0.012h_+$  &$0.004+0.028h_+$ & $-0.038$ & $-0.033$  \\ 
\hline 
\end{tabular}\vspace*{-12pt}
\end{center}
\caption{
{\small Estimates of the expectation values of the four 
charm quark current densities at origin, 
at order $\frac{\alpha_s}{m_c}$, in
$\GeV^3$. We assume $\alpha_s(m_c)\! = \!0.3$ and 
$m_c\!=\!1.15\GeV$. }}
\label{tab:esti}
\end{table}

\begin{table}[!h]
\begin{center}
\small{\begin{tabular}{|l|l|l|l|l|l|l|}
\hline
        & $\aver{B_\delta\!+\!B_d\!+\!B_f}$ & 
$\aver{E_\delta\!+\!E_d\!+\!E_f}$ & 
$\aver{2(E_\delta\!+\!E_d)\!-\!E_f}$ & 
$\aver{B_\delta\!-\!E_\delta}$ & 
$\aver{B_\delta\!+\!E_\delta}$ \\
\hline
${\rm GeV}^4$ & $0.12$  & $0.007 $ & $0.18$ & -0.03 &  
$0.03h_+$  \\ 
\hline 
$\delta \Gamma_{\rm sl}/\Gamma_{\rm sl}$
& $-0.0034$  & $5\!\cdot\!10^{-6}$  & $0.0015$ & $0.0012$ & 
$-0.00074h_+$ \\
\hline
\end{tabular}}
\caption{{\small 
Estimates of the expectation values of the density at origin
of the $D\!=\!4$ gluon field operators, in $\GeV^4$, and the
corresponding contributions to $\Gamma_{\rm sl}(b\tto c)$. 
The expectation value of $2(B_\delta\!+\!B_d)\!-\!B_f$ vanishes in the
factorization approximation.}}
\end{center}
\end{table}

%%%%%%%%%%%%
\subsubsection{$D\!=\!8$ operators at one loop}
%%%%%%%%%%

Next we estimate contributions from operators generated by integrating
out charm at one loop which does not incorporate hard gluons with
momenta $\sim\!m_c$. As discussed previously they are of order
$1/m_c^2$ and can be expected at a fraction of percent level.

Not much is known about the expectation values of higher-dimensional
operators. Usually one applies dimensional estimates with
characteristic hadronic scales entering the problems.  Although still
leaving elements of freedom, they allow to estimate the potential
magnitude of the effects. The rules for our dimensional estimates are
formulated in the second part of this section. Unfortunately it is not
possible to infer the sign of the contributions in this way. Related
to this, we would have to add various contributions -- which
proliferate for higher-dimensional operators -- incoherently assuming
they have same sign. While we consider these counting rules a useful
tool for estimating the magnitude of various terms in general, they
can quite possibly overestimate the net effect on the semileptonic
width in question.

On the other hand, we suggest a more credible way to estimate the
relevant expectation values based on the ground-state
factorization. The advantage is that it predicts the sign of all the
contributions that do not vanish in this approximation, and is free
from arbitrary factors similar to the conventional vacuum
factorization which has been used for decades with reasonable
success. It appears possible to estimate all the contributions in this
way since at one loop the effective operators appear directly in the
`color-through' form, cf.\ Eqs.~(\ref{54}), (\ref{62}), of the generic
structure
\beq
\bar{b} \pi_k \pi_\mu \pi_\nu \pi_\rho \pi_l b\,,
\label{252}
\eeq
with or without the $b$-quark spin matrix $\vec{\sigma}_b$. The axial
current ${\cal A}_j$ includes $\sigma_j$, while the 
vector current ${\cal V}$ is spin-singlet. Besides the terms with four
spacelike and one timelike derivatives coming from the chromoelectric
field, both currents contain a term with three timelike derivatives.

For each of the matrix elements with a single $\pi_0$ there is a
unique way to insert the ground state $\state{\Omega_0}$ in the
product. When $\pi_0$ is in the middle, $\nu\!=\!0$ in
Eq.~(\ref{252}), the factorization contribution vanishes due to the 
equation of motion $\pi_0 b\!=\!0$ for the $b$ field. All the other
expectation values can be evaluated using 
\beq
-\matel{\Omega_0}{\pi_k\pi_0\pi_l}{\Omega_0} = \frac{\rho_D^3}{3} 
\delta_{kl} +  \frac{\rho_{LS}^3}{6}\sigma_{kl}, \qquad
\matel{\Omega_0}{\pi_k\pi_l}{\Omega_0} = \frac{\mu_\pi^2}{3} 
\delta_{kl} -  \frac{\mu_G^2}{6}\sigma_{kl}\;.
\label{254}
\eeq
The estimates of the expectation values of the operators with three 
time derivatives follows from the SV sum rules paralleling the 
consideration of the $D\!=\!7$ operators, Eqs.~(\ref{150}-\ref{152}): 
\bea
\nonumber
-\matel{\Omega_0}{\pi_k\pi_0\pi_0\pi_0\pi_l}{\Omega_0} 
&\msp{-4} =\msp{-4}& 
\left(\tilde\varepsilon_{3/2}^2
\frac{2\rho_D^3\!-\!\rho_{LS}^3}{9} +
\tilde\varepsilon_{1/2}^2 \frac{\rho_D^3\!+\!\rho_{LS}^3}{9}\right)
\delta_{kl} \\
& +\msp{-4}&
\left(-\tilde\varepsilon_{3/2}^2 \frac{2\rho_D^3\!-\!\rho_{LS}^3}{18}
+ \tilde\varepsilon_{1/2}^2 \frac{\rho_D^3\!+\!\rho_{LS}^3}{9}\right)
\sigma_{kl},
\label{256}
\eea
where $\tilde\varepsilon^2_{3/2,\,1/2}$ are average squares of the
$P$-wave excitation energies in the $\frac{3}{2}$- and
$\frac{1}{2}$-channels, respectively. One generally has
$\tilde\varepsilon^2_{3/2,\,1/2}\!\gsim
\!\bar\varepsilon^2_{3/2,\,1/2}$ compared to the similar average
energies introduced in the previous subsection, where the lower bounds
follow from the H\"{o}lder inequality. 
Collecting everything and using the same numerical 
values as given above Eq.~(\ref{160}) and 
$\tilde\varepsilon^2_{3/2,\,1/2}=(0.4\GeV)^2$, we get 
\begin{eqnarray}
\nonumber
32\pi^2\aver{{\cal V}_0} \msp{-3}&\simeq&\msp{-4}
-\frac{2}{45 m_c^2}\left\{100\mu_\pi^2 \rho_D^3 - 
35\mu_G^2 \rho_{LS}^3-
16(\tilde\varepsilon_{3/2}^2 (2\rho_D^3\!-\!\rho_{LS}^3) +
\tilde\varepsilon_{1/2}^2 (\rho_D^3\!+\!\rho_{LS}^3))
\right\}\\[-2pt]
\label{258}
\msp{-3}&\approx &\msp{-4} -0.3\GeV^3,  \\[0.2cm]
32\pi^2\aver{\vec\sigma  \vec {\cal A}} &\simeq&
-\,\frac{2}{9 m_c^2} 
\left\{20 \mu_G^2\rho_D^3-6\mu_\pi^2\rho_{LS}^3+7\mu_G^2\rho_{LS}^3
-2[\tilde\varepsilon_{3/2}^2 (2\rho_D^3\!-\!\rho_{LS}^3)
- 2\tilde\varepsilon_{1/2}^2 (\rho_D^3\!+\!\rho_{LS}^3)]\right\}
\nonumber \\ 
\msp{-3}&\approx &\msp{-4} -0.2\GeV^3
\label{260}
\end{eqnarray}
According to our estimates the vector current dominates 
$32\pi^2 H_c^{\rm 1-loop}\!\approx\!0.3\GeV^3$, Eq.~(\ref{262});
the correction to the semileptonic width is then
\beq
\frac{\delta\Gamma_{\rm sl}}{\Gamma_{\rm
sl}}\;\rule[-12pt]{.3pt}{25pt}\raisebox{-11pt}{$\mbox{{\footnotesize 
$\,1/m_c^2\!$}}$} \approx 0.003\,.
\label{264}
\eeq

As mentioned above, we also tried educated dimensional arguments 
to estimate the higher-order operators. This was done adopting 
the following rules:

$\bullet$ The covariant derivative, either spacelike or timelike,
counts as $0.35 \GeV$, and summation over the Lorentz index brings in
the number of components. For instance, the estimate for $\mu_\pi^2$
would be $0.37\GeV^2$.

$\bullet$ Consequently, each factor of $E$ or $B$ field counts 
as $(0.35\GeV)^2$ per component.

$\bullet$ For the anticommutators of the two field strengths we put an
extra factor $\sqrt{2}$ while the commutators of field strengths or
covariant derivatives count simply as the product.

$\bullet$ The spin-dependent operators come contracted with
the $\sigma$ matrices. We treat $\vec \sigma$ as a vector with unit
components, which according to the above prescriptions reproduces 
the actual chromomagnetic expectation value.

$\bullet$ All terms in an expression obtained by integrating out
charm in a particular
current are added in moduli regardless of the sign of the
corresponding coefficient. 

Using various identities and the equation of motion for the $b$ field, 
the results for higher-order operators can often be
cast in different forms resulting in different estimates, in particular
due to the last rule or due to   
using the equation of motion for the $b$ field. 
In this case of alternative values we adopt the
smallest of the obtained numbers.

While admittedly of limited accuracy, these rules seem reasonable,
yielding the right size for the lowest-dimension expectation
values. Applying these to the estimates of $D=8$ operators in the
one-loop vector or axial current, we obtain
\beq
32\pi^2\aver{{\cal V}_0} \sim 0.15\GeV^3, \qquad 
32\pi^2\aver{\vec\sigma 
\vec{\cal A}} \sim 0.2\GeV^3,
\label{270}
\eeq
qualitatively consistent with the factorization 
estimates; the similarity for the axial current is 
probably somewhat accidental. The smaller value for the vector current
probably indicates that counting the commutator with a unit factor in
the above rules may undervalue the higher-dimension local operators.
Of course, the dimensional estimates do
not allow to specify the relative sign of the vector and the axial
contributions, and therefore would allow a larger overall
effect. Adding the two contributions in modulus yields
\beq
\frac{\delta\Gamma_{\rm sl}}{\Gamma_{\rm
sl}}\;\rule[-12pt]{.3pt}{25pt}\raisebox{-11pt}{$\mbox{{\footnotesize 
$\,1/m_c^2\!$}}$}
=
32\pi^2 \frac{(m_b-m_c)^2}{m_b^5 z_0(\frac{m_c^2}{m_b^2})}H_c \sim 
0.004 \, .
\label{193}
\eeq
In the case of the $D\!=\!7$ operators obtained at two loops we
observed significant cancellations between different terms in the
expressions for a particular current. This suggests that the 
rule of incoherently adding all contributions may noticeably 
overestimates the magnitude of the actual expectation
values. That is what we found in estimating the matrix elements
through the ground-state factorization, although here the difference 
is not large numerically due to dominance of the axial current
contribution.

%%%%%%%%%%%%
\subsection{Exponential effects}
%%%%%%%%%%
 
For a sufficiently heavy quark, in our case charm, the leading
operators in the power expansion when integrating it out must
accurately describe the nonperturbative effects associated
with these fields.
This expansion -- like all such power series expansions -- is as a
rule only asymptotic. 
Therefore it would leave out generic exponential terms scaling
like $e^{- 2m_c/\mhad}$; the exponent, in general, could even be a
non-integer power of $m_c$. For a dedicated discussion we refer the
reader to the papers \cite{inst,shifioffe}. A related consideration 
actually suggests that twice the charm quark mass enters the
exponent. 

It is possible to develop an approach similar to that of
Ref.~\cite{inst} to analyze the asymptotic behavior of the exponential
contributions or to estimate their effect in a concrete dynamical
ansatz for the nonperturbative light fields. We did not attempt this
here. Nevertheless, charm quarks can be considered as marginally heavy
in actual QCD. Virtual charm effects must definitely be suppressed,
but not necessarily to the extent where the asymptotic regime can be
trusted. They are expected to decrease with growing $m_c$. Hence, a
safe upper bound would come from similar effects of the nonvalence $d$
quark in the semileptonic $b\tto u\,\ell\nu$ decays of $B^0$
mesons. The latter can potentially be at the scale of a couple percent
\cite{vub}. Assuming the charm mass suppression to be a factor of five
or stronger, and accounting for the phase space difference, we arrive
at the scale
\beq
H_c \sim 0.001 \GeV^3\;,
\label{198}
\eeq
or $\delta \Gamma_{\rm sl}/ \Gamma_{\rm sl} \!\sim\! 0.4\%$. Whether
the truncated power expansion yields such terms, or results in smaller
effects due to specific cancellations, we do not consider it justified
to {\it a priori} rule out such effects up to the $0.5\%$ level.

Therefore, the most direct way is to introduce the soft-charm effects 
as required by the OPE, yet described by expectation values taken 
as free parameters and 
analyze the observable effects they induce. Then
one can obtain direct experimental bounds from precision measurements
of the inclusive observables in $B\tto X_c \ell\nu$ decays, thereby
placing more reliable upper bounds on the limitations they
impose on extraction of $V_{cb}$. 

%%%%%%%%%%%%
\section{Phenomenology}
%%%%%%%%%%%%

All the inclusive semileptonic $B$ decay distributions are described 
in terms of five structure functions $w_i(q^2,q_0)$ 
representing different Lorentz combinations \cite{koy}.
In the $1/m_b$ expansion through the OPE the effects of the
four-quark charm operators appear in the structure functions
as a delta-function at maximal
$q^2\!=\!(m_b\!-\!m_c)^2$ and $q_0\!=\!m_b\!-\!m_c$, 
$$
\delta w_i \propto \delta^4(q_\mu\!-\!(m_b\!-\!m_c)v_\mu)\propto
\delta(q^2\!-\!(m_b\!-\!m_c)^2)\:
\frac{1}{\sqrt{q_0^2\!-\!q^2}}\delta(q_0\!-\!(m_b\!-\!m_c))\;.
$$
The strong interactions smear out this distribution over the range 
$\sim \!\mhad \!\sim\!\Lam$ in $\sqrt{q^2}$ and in $q_0$. However in the
fully inclusive characteristics this is an effect of a higher order in
$1/m_b$, and we neglect it here. Likewise, we disregard the
perturbatively-induced tails towards lower values of $q^2$ and $q_0$,
associated with the anomalous dimension of the four-quark operators
\cite{WA}.

Of the five weak decay structure functions of $B$ meson 
three contribute to the 
decays with massless leptons,
\bea
\nonumber
\frac{{\rm d}^3 \Gamma}{{\rm d}E_{\ell\,}  {\rm d}q^{2\,} {\rm d}q_0 }
&\msp{-4}= \msp{-4}& \frac{G_F^2 |V_{cb}|^2}{32\pi^4}\,
\theta\!\left(q_0\!-\!E_\ell\!-\!\mbox{$\frac{q^2}{4E_\ell}$}\right)
\theta(E_\ell) 
\,\theta(q^2) \;\times \\ 
& & \msp{20} \left\{
2 q^2 w_1+[4E_\ell (q_0\!-\!E_\ell)\!-\!q^2]w_2 +
2q^2(2E_\ell\!-\!q_0) w_3
\right\} .
\label{320}
\eea
Yet only two of five Lorentz structures, 
$-\delta_{\mu\nu}$ and  $v_\mu v_\nu$ are
independent at the point where $q^2\!=\!q_0^2$ (i.e., $\vec
q^{\,2}\tto 0$). These are described by the structure functions $w_1$
and $w_2$, respectively. Contributions from $w_{3,4,5}$ either vanish
being proportional to $\vec q$ or reduce to that of $w_{1,2}$. Two 
kinematic step-functions in Eq.~(\ref{320}) effectively yield a
$\delta(E_\ell-\frac{m_b-m_c}{2})$-type lepton spectrum, as expected
(it is also smeared in reality over an interval
$\sim\!\mhad$). Furthermore, for massless leptons the effect of $w_2$
vanishes as well, since one has
\beq
\Gamma_{\rm sl} \!=\! \frac{G_F^2 |V_{cb}|^2}{16\pi^4}
\int_{0}^{m_b^2} \!{\rm d}q^2   \int_{q_0>\sqrt{q^2}}^{m_b} {\rm d}q_0
\sqrt{q_0^2\!-\!q^2} \left(q^2 w_1(q^2\!,q_0) 
+\frac{1}{3}(q_0^2\!-\!q^2) w_2(q^2\!,q_0) \right)\;.
\label{322}
\eeq
Therefore, for the correction to the total width in our case we have
\beq
\delta_\chi\Gamma_{\rm sl} \!=\! \frac{G_F^2 |V_{cb}|^2}{16\pi^4}
\int\!{\rm d}q^2   \int {\rm d}q_0\:q^2
\sqrt{q_0^2\!-\!q^2} \,\delta_\chi w_1(q^2\!,q_0) 
\;,
\label{324}
\eeq
with
\beq
\delta_\chi w_1(q^2\!,q_0) \!\simeq\!\frac{8\pi^3}{3} H_c \,
\delta(q^2\!-\!(m_b\!-\!m_c)^2)\:
\frac{1}{\sqrt{q_0^2\!-\!q^2}}\delta(q_0\!-\!(m_b\!-\!m_c))\;.
\label{326}
\eeq
Eqs.~(\ref{320}), (\ref{324}), (\ref{326}) allow one to evaluate the
soft-charm effects in various inclusive moments used to experimentally
determine the heavy quark parameters, in terms of $H_c$. 

The dimensionless ratio 
$$
\chi=\frac{\delta_\chi\Gamma_{\rm sl}}{\Gamma_{\rm sl}}\simeq 
32\pi^2 \frac{(m_b-m_c)^2}{m_b^5 z_0(\frac{m_c^2}{m_b^2})} \, H_c \simeq
3 \GeV^{-3} \cdot H_c
$$
determining the correction to the total semileptonic width, sets the
overall scale of the effects. According to our estimates, it can
hardly exceed the one percent level, which places the first benchmark for
the accuracy required to detect or constrain them further. Since 
$\chi\!\ll\!1$, in what follows we will keep only the terms linear in
$\chi$ and drop everything that scales like $\chi^2$ or higher. 

The average charged lepton energy receives a correction proportional to
$\chi$:
\beq
\delta_\chi \aver{E_\ell} \!\simeq\!
R_0^{-1}\,\chi\left(\mbox{$\frac{m_b\!-\!m_c}{2}$}-\aver{E_\ell}_0\right)\;,
\label{340}
\eeq
where $\aver{E_\ell}_0\simeq 1.38\GeV$ is the conventional average
without the soft-charm effects; it is strongly dominated by the parton-level
piece. ($R_0$ is the decay fraction left by applying kinematic cuts;
$R_0\!=\!1$ for the full moment.)
Since the related excess or depletion of the rates emerges
near $E_\chi\!\simeq\!\frac{M_B-M_{D^*}}{2}\simeq 1.64\GeV$ which is
close to the bare lepton average $\aver{E_\ell}_0$, the impact of the
charm field expectation values 
on $\aver{E_\ell}$ is strongly suppressed. In principle,
$\aver{E_\ell}_0$ grows when a lower cut on $E_\ell$ is applied,
while the soft-charm contribution remains unaltered up to $E_{\rm
cut}^\ell\!\simeq\!1.3\GeV$. However, the theoretical precision is
probably insufficient to detect the mismatch in the overall
normalization of the low-$E_\ell$ part of the spectrum at the
sub-percent level. 

A similar problem would plague the utility of the higher lepton
energy moments:
\bea
\nonumber
\delta_\chi \aver{[E_\ell\!-\!\aver{E_\ell}]^2} \msp{-4} &\simeq &
\msp{-4}
R_0^{-1}\,\chi\left((\mbox{$\frac{m_b\!-\!m_c}{2}$}-\aver{E_\ell}_0)^2-
\aver{[E_\ell\!-\!\aver{E_\ell}]^2}_0\right)\\
\delta_\chi \aver{[E_\ell\!-\!\aver{E_\ell}]^3} \msp{-4} &\simeq &
\msp{-4}
R_0^{-1}\,\chi\left((\mbox{$\frac{m_b\!-\!m_c}{2}$}-\aver{E_\ell}_0)^3-
\aver{[E_\ell\!-\!\aver{E_\ell}]^3}_0\right)
\;;
\label{346}
\eea
the numerical values of the moments with the subscript zero are
measured in a number of experiments and/or are calculated theoretically,
see, e.g.\ Ref.~\cite{slcm}. The underlying feature common 
to all lepton moments is that they are dominated by the
tree-level or `partonic' contribution, with all nonperturbative effects being
small corrections. At the same time the soft-charm effect is not enhanced,
but rather suppressed by the smallness of the difference between
$E_\chi$ and $\aver{E_0}$.

At first glance, the moments of the hadronic invariant mass squared
may look better since most of  the 
deviation of $\aver{M_X^2}$ from $M_{D^*}^2$
is due to nonperturbative effects, so that the `parton background' 
is suppressed. 
However, the contributions we focus on are expected
to populate just the domain of $M_X$ near $M_D$ or slightly above,
where the bulk of the decays happen:
\bea
\nonumber
\delta_\chi \aver{M_X^2} \msp{-4} &\simeq & \msp{-4}
R_0^{-1}\,\chi\left(M_{D^*}^2-\aver{M_X^2}_0\right)\\
\nonumber
\delta_\chi \aver{[M_X^2\!-\!\aver{M_X^2}]^2} \msp{-4} &\simeq & \msp{-4}
R_0^{-1}\,\chi\left((M_{D^*}^2-\aver{M_X^2}_0)^2-
\aver{[M_X^2\!-\!\aver{M_X^2}]^2}_0\right)\\
\delta_\chi \aver{[M_X^2\!-\!\aver{M_X^2}]^3} \msp{-4} &\simeq & \msp{-4}
R_0^{-1}\,\chi\left((M_{D^*}^2-\aver{M_X^2}_0)^3-
\aver{[M_X^2\!-\!\aver{M_X^2}]^3}_0\right)
\;.
\label{350}
\eea
Detecting such shifts would require controlling theory predictions for
the moments, roughly speaking, at a percent level, which does not look
realistic, see, e.g.\ Ref.~\cite{slcm}. (For $\aver{M_X^2}$ this
would apply to $\aver{M_X^2}_0 \!-\! M_{D^*}^2 \!\simeq\!0.6\GeV^2$.)

A realistic way to look for the nonperturbative 
soft-charm effects is studying directly the
$q^2$-distribution, or the associated $q^2$ moments. By their nature
the contributions in question are located around the maximal $q^2
\!\simeq\!(M_B \!-\!M_{D^*})^2\!\simeq\!10.7\GeV^2$, while the
average $q^2$ is less than half of this maximal
value. Therefore, the $q^2$ moments appear to be sensitive:
\bea
\nonumber
\delta_\chi \aver{q^2} \msp{-4} &\simeq & \msp{-4}
R_0^{-1}\,\chi\left((M_B\!-\!M_{D^*})^2-\aver{q^2}_0\right)\\
\nonumber
\delta_\chi \aver{[q^2\!-\!\aver{q^2}]^2} \msp{-4} &\simeq & \msp{-4}
R_0^{-1}\,\chi\left([(M_B\!-\!M_{D^*})^2-\aver{q^2}_0]^2-
\aver{[q^2\!-\!\aver{q^2}]^2}_0\right)\\
\delta_\chi \aver{[q^2\!-\!\aver{q^2}]^3} \msp{-4} &\simeq & \msp{-4}
R_0^{-1}\,\chi\left([(M_B\!-\!M_{D^*})^2-\aver{q^2}_0]^3-
\aver{[q^2\!-\!\aver{q^2}]^3}_0\right)
\;,
\label{356}
\eea
where $\aver{q^2}_0\!\approx\!4.3\GeV^2$,
$\aver{[q^2\!-\!\aver{q^2}]^2}_0\!\approx\!7.5\GeV^4$ and
$\aver{[q^2\!-\!\aver{q^2}]^3}_0\!\approx\!7\GeV^6$ (without lower cut
on $E_\ell$). Similarly, one can study the moments of $q_0$ instead of
the moments of $q^2$. 

In practice, precision measurements of inclusive $B$ decays usually
require a lower cut on the charge lepton energy around $1\GeV$. To
facilitate comparison of experimental data with theory, 
we present here the numerical predictions for the $q^2$ moments, which 
are available applying the OPE \cite{slcm,slsf,sfpert}.\footnote{We
give here the results obtained with the same simplifications as in 
Ref.~\cite{slcm} which proved to be a good approximation. The complete
cut-dependence in the perturbative corrections is available 
\cite{slsf,sfpert}; therefore, the estimates can be easily refined in
this respect, if necessary.} For instance,
based on the recent fit \cite{buch} to the existing data on $B$
decays, we get for the lepton energy cut at $1\GeV$ 
\bea
\nonumber
\aver{q^2} \msp{-4} &\simeq & \msp{-4}
4.83\GeV^2 +7.2\GeV^2 \,\chi\,,  \\ 
\nonumber
\aver{[q^2\!-\!\aver{q^2}]^2} \msp{-4} &\simeq & \msp{-4}
7.53\GeV^4 +33\GeV^4 \chi \,,\\ 
\aver{[q^2\!-\!\aver{q^2}]^3} \msp{-4} &\simeq & \msp{-4}
4.6\GeV^6 +245\GeV^6 \chi\,. 
\label{360}
\eea 
The dependence on the usual heavy quark parameters can be approximated
in the way similar to the one adopted in Ref.~\cite{slcm}:
{\small\bea
\nonumber
{\cal M}(m_b,m_c,\mu_\pi^2,\mu_G^2, \rho_D^3,
\rho_{LS}^3;\alpha_s) \msp{-4}&=&\msp{-4}
V + B\,(m_b\!-\!4.6\GeV)+C\,(m_c\!-\!1.15\GeV) \\
&&\msp{-.9}+ P\,(\mu_\pi^2\!-\!0.4\GeV^2)
+D\,(\rho_D^3\!-\!0.2\GeV^3) \qquad 
\label{410}
\\
\nonumber
&&\msp{-.9}+  G\,(\mu_G^2\!-\!0.35\GeV^2)
+L\,(\rho_{LS}^3\!+\!0.15\GeV^3)
\;.
\eea}
\vspace*{-5.5mm}

\noindent
The reference values $V$ and the
linear dependence coefficients $B$ to $L$ are given in Tables 3 to 5. 
$V$ has dimension of the moment ${\cal M}$ itself
and the quoted numbers are in $\GeV$ to the corresponding power; the values
of the coefficients $B$ to $L$ are likewise in
the proper power of $\GeV$, and $E_{\rm cut}$ are shown in $\GeV$.
We note that the preliminary data reported by CLEO are in agreement
with theory predictions without significant nonperturbative charm 
effects. Moreover, an estimated accuracy of $\pm 0.1\GeV^2$ in 
$\aver{q^2}$ and  $\pm 0.5\GeV^4$ 
in $\aver{[q^2\!-\!\aver{q^2}]^2}$ translates into the sensitivity in
$\chi$ of $\,0.015$. 
At the percent level of
precision the reliability of theory predictions becomes important.
Matching the demand would require calculating the $\alpha_s$-corrections 
to the Wilson coefficients of the power-suppressed operators, 
which is presently the limiting factor. 
\vspace*{7mm}\\
\begin{tabular}{|c|l|l|l|l|l|l|l|}\hline
$ E_{\rm cut}  $& ~\hfill $ V $\hfill~ & \hfill~ $B$ \hfill~ & ~\hfill
$ C $ \hfill~
& ~\hfill $P$ \hfill~&~\hfill $D$ \hfill~&~\hfill $G$ 
\hfill~&\hfill~ $L $ \hfill~ \\ \hline
$ 0 $&$ 4.31992 $&$ 2.6176 $&$-2.1118 $&$-0.008429   $&$-1.138 $&$
 -0.627 $&$  0.1361 
$ \\ \hline
$ 0.5 $&$4.41171  $&$2.6084 $&$ -2.1106 $&$-0.008009 $&$ -1.147
$&$-0.6301 $&$ 0.1369  $\\ \hline
$ 1.0 $&$4.78331 $&$2.6724 $&$ -2.2266 $&$ -0.02488 $&$-1.311 $&$
-0.709 $&$  0.146       $\\ \hline
$ 1.5 $&$5.13019 $&$3.3376 $&$-2.9504 $&$-0.1478  $&$  -2. $&$
-1.135 $&$ 0.1559  $\\ \hline
\end{tabular}\vspace*{2mm}\\
Table 3.~~{\small First moment of the lepton pair
invariant mass $\aver{q^2}$.}
\vspace*{7mm}

\noindent
\begin{tabular}{|c|l|l|l|l|l|l|l|}\hline
$ E_{\rm cut}  $& ~\hfill $ V $\hfill~ & \hfill~ $B$ \hfill~ & ~\hfill
$ C $ \hfill~
& ~\hfill $P$ \hfill~&~\hfill $D$ \hfill~&~\hfill $G$ 
\hfill~&~\hfill $L$ \hfill~ \\ \hline
$ 0 $&$7.23244  $&$ 9.3514 $&$ -7.2682 $&$ -0.04725 $&$ -7.737  
$&$ -3.167 $&$ 0.6866   $\\ \hline
$ 0.5 $&$ 7.1329  $&$ 9.2728 $&$-7.182  $&$ -0.04353 $&$ -7.773 $&$
-3.154 $&$ 0.686   $\\ \hline
$ 1.0 $&$ 7.17852$&$  9.1852$&$  -6.982$&$  -0.0397 $&$ -8.393$&$ 
 -3.233$&$  0.7099  $\\ \hline
$ 1.5 $&$ 6.95888 $&$  10.863 $&$  -7.9839 $&$  -0.3896 $&$ 
 -12.29 $&$  -4.649 $&$  0.7491   $\\ \hline
\end{tabular}\vspace*{2mm}\\
Table 4.~~{\small Second $q^2$-moment  
with respect to average, 
$\aver{(q^2\!-\!\aver{q^2})^2}$}\vspace*{7mm}

More detailed kinematic measurements allowing to study the dependence
of the integrated rates upon correlated changes in the limits on $q^2$
and $q^0$, can potentially further improve the sensitivity to the 
`soft charm' 
effects by more effectively emphasizing the kinematics where the
corresponding physics resides. A certain possibility along these lines
has been discussed in Ref.~\cite{imprec}.
\vspace*{7mm}

\noindent
\begin{tabular}{|c|l|l|l|l|l|l|l|}\hline
$ E_{\rm cut}  $& ~\hfill $ V $\hfill~ & \hfill~ $B$ \hfill~ & ~\hfill
$ C $ \hfill~
& ~\hfill $P$ \hfill~&~\hfill $D$ \hfill~&~\hfill $G$ 
\hfill~&\hfill~ $L $ \hfill~ \\ \hline
$ 0 $&$ 6.92486 $&$  16.488 $&$  -10.568 $&$  -0.2179 $&$ 
 -44.39 $&$  -12.43 $&$  2.683  $\\ \hline
$ 0.5 $&$ 6.11989 $&$  15.69 $&$  -9.8461 $&$  -0.2234 $&$  
-44.02  $&$ -12.22  $&$ 2.634  $\\ \hline
$ 1.0 $&$ 2.60892  $&$ 10.978  $&$ -4.9345 $&$  -0.0397 $&$  
-44.1  $&$ -11.35 $&$  2.536   $\\ \hline
$ 1.5 $&$ -2.03212  $&$ 7.5062 $&$  2.9761 $&$  -0.9282  $&$ -62.41  
$&$ -16.01  $&$ 2.829  $\\ \hline
\end{tabular}\vspace*{2mm}\\
Table 5.~~{\small Third $q^2$-moment 
with respect to average, 
$\aver{(q^2\!-\!\aver{q^2})^3}$}\vspace*{1.4mm}\\

Ultimately, the analysis of the inclusive semileptonic decay data
should include the value of $H_c$ in the fit, along with the other
heavy quark parameters as required by the OPE. A simplified procedure
would be using the heavy quark parameters currently extracted from the
lepton energy and hadronic mass moments ignoring these effects, to
compute the $q^2$ moments. Comparing those with the measured ones
allows to infer bounds on $H_c$ and in this way probe the
nonperturbative charm contributions, or even detect them.

\section{Discussions and conclusions}

With $|V_{cb}|$ and other heavy quark parameters like $m_b$ and $m_c$
having been extracted from $B \to l \nu X_c$ with high precision, 
scrutinizing even small contributions comes onto the agenda. 
In this paper we have analyzed the effects of the
nonperturbative QCD interactions of charm quarks in inclusive decays
of $B$ hadrons, specifically semileptonic $b\tto c\,\ell \nu$ 
transitions. These effects
had so far not been included in practical applications of the OPE;
allowance for them was made in Ref.~\cite{imprec} in assessing the
theoretical uncertainty.

Inclusive weak decays of heavy flavor hadrons admit a local OPE, which
is crucial for our analysis. It allows us to study nonperturbative
charm effects in a model-independent manner. 
We have shown that these
contributions have a well-defined meaning in the OPE described by
expectation values of four-heavy-quark  operators like 
$\matel{B}{\bar{b} \Gamma c \bar{c} \Gamma b}{B}$ (or similar
operators with additional derivatives once terms of higher order in
$1/m_b$ are considered). 
The OPE actually
requires the inclusion of these four-heavy-quark operator terms,
and they cover the impact of the soft-charm dynamics on $B$ decays without
double- or undercounting. For inclusive widths they scale with $m_b$ 
like $1/m_b^3$. 

The expectation values $\matel{B}{\bar{b} \Gamma c \bar{c} \Gamma
b}{B}$ can be evaluated, through an expansion in powers of $1/m_c$, in
terms of the higher-dimensional `usual' heavy quark operators, i.e.\
those without charm quarks. Ignoring radiative QCD effects we obtain
terms scaling like $\frac{\Lam^2}{m_c^2}$ for an overall contribution
$\propto \frac{\Lam^5}{m_c^2 m_b^3}$. Yet once hard gluons are
considered, there appear contributions in the rate which fade out only
as the first power of $1/m_c$, $\alpha_s(m_c)\frac{\Lam^4}{m_c
m_b^3}$.  In the picture where nonperturbative charm emerges through
the higher Fock states in the $B$ meson wavefunction (so-called
`Intrinsic Charm', see below) these terms describe the interference of
the effects with and without a $c\bar{c}$ pair made possible through
hard gluon annihilation.

For a numerical evaluation of the nonperturbative charm effects we
have elaborated a method for estimating the higher-dimensional $b$
quark expectation values; it can be used in other applications of the
heavy quark expansion as well. We also found a strong enhancement of
the two-loop Wilson coefficients for the $D\!=\!7$ operators (i.e.,
${\cal O}\left( \frac{\alpha_s(m_c)}{m_c^1}\right)$ effects) evidently
related to the peculiarity of the Coulomb interaction of static
quarks. It could potentially lead to the dominance of the ${\cal
O}\left( \frac{\alpha_s}{m_c}\right)$ contributions with individual
ones as large as half a percent. However, we found significant
cancellations between different contributions, see Table~2. In view of
the approximations in both calculating the Wilson coefficients and in
estimating the corresponding expectation values, we cannot take the
resulting sub-permill numbers at face value. However, a net
contribution exceeding $0.005\,\Gamma_{\rm sl}$ looks improbable.

As a result of the above cancellations an appreciable effect might be
expected from $D\!=\!8$ operators with the Wilson coefficients
generated at one loop. The natural scale of these contributions can be
a few permill in $\Gamma_{\rm sl}$; explicit calculations led us to
expect an overall effect of about $0.003\,\Gamma_{\rm sl}$. Even
considering the approximate nature of our estimates, we conclude that
the effects associated with nonperturbative charm dynamics should not
downgrade the accuracy in extracting $V_{cb}$ at a percent level as
long as no $1/m_c$ expansion is unnecessarily built into the analysis
of the semileptonic data.

In addition, we have suggested a more or less direct way to
experimentally probe these effects without relying on estimates of
the $\bar{b}b\,\bar{c}c$  expectation values in a $1/m_c$ expansion. 
Such experimental tests are sensitive, of course, to the total 
expectation values, including possible exponential
contributions $\propto \! e^{-2m_c/\mhad}$ which would be left out in
their $1/m_c$ expansion. The $q^2$-distributions like the higher 
$q^2$-moments are particularly constraining in this
respect. With the present experimental capabilities it should be
possible to constrain -- or measure -- these effects to the level
relevant for the precision in $V_{cb}$ down to a fraction of one
percent.

It has been forcefully argued in many papers over the years that  
hadrons may contain $c \bar c$ pairs as a higher Fock component in their
wave function. This particular mechanism is usually referred to as
`Intrinsic Charm' (IC). 
More specifically it has been suggested \cite{evade}
that such a component for $B$ mesons could vitiate the strong CKM
hierarchy in their decay modes expected when only `valence' (and
`light-sea') quarks are included in the wavefunction. This 
`Intrinsic Charm' picture can be used to visualize many of the effects
we have discussed in this paper. However the relationship between 
the soft-charm effects
discussed here and the `Intrinsic Charm' ansatz is that of an analogy
and an illustration rather than a genuine connection. At the
technical level, the
existence of a full-fledged OPE treatment is crucial for our analysis,
and that is available for inclusive, yet not exclusive transitions. The
intrinsic charm picture, on the other hand, taken literally would seem
to apply to any type of process, and actually has mostly been
discussed in the context of exclusive channels.

Even for inclusive decays the connection is tenuous at best. The
correction to $\Gamma_{\rm sl}(b\tto c)$ cannot be directly related to
the admixture of $c\bar{c}$ pairs in the $B$ meson wavefunction. The
presence of virtual $c\bar{c}$ in the initial state does not
necessarily change in a definite way the decay rate or other fully
inclusive characteristics, much in the same way as the presence of the
spectator quark per se does not affect the decay rate of the heavy
flavor hadrons \cite{buv}. Charm quarks interacting with the gluon
background potentially shape, to some extent, the structure of the $B$
meson wavefunction and, hence, the usual nonperturbative parameters
$\mu_\pi^2$, $\mu_G^2$, ..., which affects the rates -- yet the
interaction of light quarks, both sea and valence, has conceptually
the same influences. Therefore, this type of effect is not specific to
`Intrinsic Charm'. 

As discussed before, without perturbative corrections the 
leading nonperturbative charm expectation values are suppressed as
$\frac{\Lam^2}{m_c^2}$; i.e.\ the soft-charm correction to the width scales 
like $ \frac{\Lam^5}{m_c^2m_b^3}$. The $1/m_c^2$-dependence 
looks consistent with a $c\bar{c}$ Fock state admixture in the $B$
wave function $\propto \!\frac{\Lam}{m_c}$ which leads to a term 
${\cal O}\left( \frac{\Lam^2}{m_c^2}\right)$ for the $c\bar{c}$ pair
probability. However, hard gluon corrections give rise to contributions
$\propto \alpha_s(m_c)\frac{\Lam^4}{m_c m_b^3}$ without an analogue in
the naive probabilistic `Intrinsic Charm' picture.

Lastly, `Intrinsic Charm' is meant to represent an
effect due to an initial state configuration. The OPE, on the other
hand, by necessity combines initial and final state effects and often does
not allow a clear separation between the two. The soft extra 
$c\bar{c}$ pair may affect the decay rate showing up in the initial 
or in the final state, or without appearing at all -- the OPE
determines the aggregate effect of all mechanisms, including simply 
the nonperturbative corrections to the
propagation of the final state quark produced in the decay.

A physical interpretation of the four-quark expectation
values with both $b$ and $c$ becomes transparent if one adopts 
a naive factorization
approximation in the usual IC ansatz: they would correspond to the $\bar{c}$
density {\sf at the origin}, $|\Psi _{\bar{c}}(0)|^2$, rather than the
overall $c\bar{c}$-pair probability integrated over the $B$ meson
volume: 
\beq
w_{c\bar{c}}= \int {\rm d}^3{\bf x}\: |\Psi _{\bar{c}}({\bf x})|^2\,.
\label{510}
\eeq
In this equation we have tacitly assumed that only one pair of $c\bar{c}$
can be present, and that each charm antiquark in the wavefunction 
must be accompanied by one charm quark. The presence of color and 
spin degrees of freedom was also ignored. 
The four-quark expectation values
actually select the particular projections in those spaces.

Evidently there can be no direct
relation between the admixture of a $c\bar{c}$ component in the
wavefunction $w_{c\bar{c}}$ and the $\bar{c}$ density at origin
$|\Psi_{\bar{c}}(0)|^2$ corresponding to the four-quark expectation
values. Yet a general dimensional estimate
\beq
|\Psi_{\bar{c}}(0)|^2 \approx \frac{1}{\frac{4}{3}\pi R_0^3}\,
w_{c\bar{c}}
\label{512}
\eeq
involving the effective size $R_0$ of the $B$ meson bound state 
should provide some idea about the possible scale in relating the
two quantities. They differ in dimension by mass to the third power,
the factor which goes into the $1/m_b^3$ suppression of the
soft-charm effects in the inclusive decay rates. The generic 
`Intrinsic Charm' contributions do not have to be suppressed by
powers of $1/m_b$. According to the estimate paralleling 
Eqs.~(\ref{176})-(\ref{181}), one has 
\beq
\frac{4}{3}\pi R_0^3 \approx
\left(\frac{M^2}{4\pi}\right)^{-\frac{3}{2}}\approx (0.008\GeV^3)^{-1}
\label{514}
\eeq
Obviously relations of this sort cannot account 
for the dynamic details like the Coulomb enhancement we have observed in
Sect.~2. 

Theory has much less to say rigorously about exclusive modes, where an
`Intrinsic Charm' component may naturally have a bigger
impact. There arises even a fundamental concern inhibiting immediate
answers: Does `Intrinsic Charm' really represent a genuinely
independent contribution to decay amplitudes when 
embedding parton model effects into real QCD,  or may it in some
instances involve
double counting? This looks especially nontrivial in the context of a
quantum field theory, where the highly virtual states are usually
absorbed into various renormalization factors. Since the charm pairs
in a static hadron have to be viewed as highly virtual, it may seem to
be the case. The situation is additionally aggravated by the fact
that the originally introduced ``Penguins'' as particular
short-distance-induced local operators
\cite{penguins} are often used now in a rather loose and wide sense. In
particular, ``generic Penguin'' diagrams characterized only by their
topology, can often be deformed in such a
way as to seemingly exemplify the IC effects, as illustrated, for
instance, by Figs.~1. In fact, a generic diagram for the
nonperturbative correction to the charm propagation in the $b\tto c$
decays similarly includes contributions which can be interpreted 
as an IC effect with subsequent $\bar{c}$ annihilation. At the
same time, it was shown \cite{mirage} that the rescattering processes 
can formally $1/m_b$-dominate over the bare annihilation-type
contributions; this applies to the charm quark annihilation as well
at sufficiently large $m_b$.

\renewcommand{\baselinestretch}{0.6}

\thispagestyle{plain}
\begin{figure}[hhh]\vspace*{-4mm}
\begin{center}
\mbox{\epsfig{file=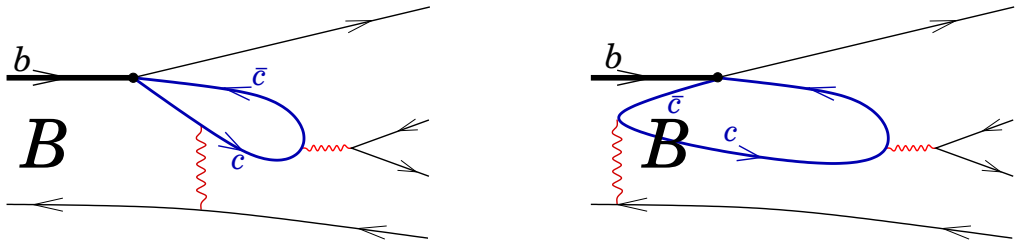,width=135mm}}
\end{center}
\vspace*{-4.0mm}
\caption{  
{\small A diagram for a `generic Penguin' process (left) where the decay along
the channel $b\tto c\bar{c}\,q$ produces a final state without hidden
charm, simultaneously includes the `Intrinsic Charm'-type contribution to
the same final state (right) if both `intrinsic' $c$ and $\bar{c}$
annihilate with the quarks produced in the decay.
}
}
\end{figure}

\renewcommand{\baselinestretch}{1}

Without going into further details here, we note that an essential
stumbling block in clarifying these principal elements here is lack,
as of now, of a consistent OPE procedure which could be applied to
such exclusive decay processes in Minkowski space. For instance, a
clear separation is missing of which effects are absorbed into the
decay matrix elements and which corrections are attributed to the effective
weak decay Hamiltonian. Related to this, there is no clear
understanding of how to integrate out the high-frequency/high-mass modes
and what type of an effective Minkowskian field theory for exclusive
decays would result from this.  A more complex answer here may, in
principle, allow for more significant strong-interaction corrections
related to non-valence charm quarks, which would be absent for
much heavier charm quarks.
\vspace*{2mm}

\noindent
{\bf Acknowledgments:} We greatly benefited from many conversations
with S.~Brodsky, S.~Gardner, Th.~Mannel, M.~Shifman, A.~Vainshtein 
and M.~Voloshin. We are grateful to S.~Trine for directing our
attention to paper \cite{trine}, which helped us to identify an
inconsistency in one of our transformation routines.
N.U., I.B.\ and R.Z.\ gratefully acknowledge the hospitality of TPI,
Lab.\ de Physique Th\'{e}orique of the Universit\'{e} de 
Paris Sud and of the Institute for Theoretical Physics 
of the University of Z\"urich, respectively, where 
a part of this study was done.
R.Z.\ was partly supported by the Swiss National Science Foundation and by
the EU-RTN Programme, Contract No.\ HPEN-CT-2002-00311, ``EURIDICE''.
This work was supported  by the NSF under grant number PHY-0355098.

\appendix

\section*{Appendix}
\setcounter{equation}{0}
\renewcommand{\theequation}{A.\arabic{equation}}
\renewcommand{\thetable}{\Alph{table}}
\setcounter{section}{0}
\setcounter{table}{0}

\section{Conventions}
\label{app:conv}
\subsubsection*{Gauge fields}

The ordinary and covariant derivatives read
\begin{equation}
\partial_\alpha = \frac{\partial}{\partial x^\alpha}\,, \quad \quad  
D_\alpha = \partial_\alpha -i t^a A^a\,,
\end{equation}
with $t^a = \lambda^a/2$ and $\lambda^a$ the usual hermitian 
$SU(3)$ Gell-Mann matrices
\begin{equation}
[\lambda^a,\lambda^b] = 2if^{abc}\lambda^c\,, 
\quad \quad {\rm Tr} [\lambda^a\lambda^b] = 2 \delta^{ab}\;.
\end{equation}
The field strength tensor and its dual are defined as  
\begin{equation}
G_{\alpha\beta} = G_{\al\be}^a t^a = i[D_\al,D_\be] \; , \; 
\tilde G_{\al\be} =
\frac{1}{2}\epsilon_{\al\be\gamma\delta}G^{\gamma\delta} \; . 
\end{equation}

\subsubsection*{Electromagnetic notations}

In terms of the field strength tensor the electromagnetic fields read 
\begin{eqnarray}
E_i &=& F_{0i} \\
B_i &=& \mbox{$\frac{1}{2}$}\epsilon_{ijk}F_{jk} \quad (F_{ij} = 
\epsilon_{ijk}B_k) \quad .
\end{eqnarray}
The square of the field strength tensor is 
\begin{equation}
F_{\mu\nu}F^{\mu\nu} = 2(B^2\!-\!E^2) \; .
\end{equation}
Greek indices $\alpha$ run from $0$ to $3$, Latin indices $a$ run from
$1$ to $3$, and the summation
convention is to be understood in the following way:
\begin{equation}
a \cdot b = a_\mu b^\mu = 
a_0 b_0 - a_1 b_1 -a_2 b_2 -a_3 b_3 = a_0 b_0 - a_k b_k\;.
\end{equation}
\subsubsection*{The $\epsilon$-tensor}
We use the $\epsilon$-tensor convention from \cite{mpolyak} which is 
the same as in Itzykson \& 
Zuber as opposed to Bjorken \& Drell,
\begin{equation}
\epsilon^{0123} = -\epsilon_{0123} = 1\;.
\end{equation}

\section{Currents in the Fock-Schwinger gauge}
\label{app:1loop}

Here we sketch the technique used for
calculating the charm currents in the external gluon field.
We use the Fock-Schwinger gauge and work in momentum space.
Our master equation is 
\begin{equation}
\label{eq:qed}
\aver{\bar{c}^a \Gamma  c^b(0)}_A = i
\int \frac{d^4k}{(2\pi)^4} 
{\rm Tr}_\gamma[ \Gamma \frac{1}{\s{\tilde{k}}\!-\!m_c} ]_{ba} \cdot 1 \; ,
\end{equation} 
where $\Gamma$ is a matrix in Dirac space, $a,b$ are 
color indices and the `propagator' must be understood as an operator:
\begin{equation}
(\tilde{k}_{\mu})_{ab} =
k_{\mu}\delta_{ab}-\frac{i}{2}(G_{\al\mu})_{ab}(0) 
\partial_{k^{\al}} 
-\frac{1}{3}
(D_\lambda G_{\al\mu})_{ab}(0) 
\partial_{k^{\al}}\partial_{k^{\lambda}} + \ldots,
\label{a11}
\end{equation}
with $\partial_{k_\al} = \frac{\partial}{\partial k^\al}$. We expand
$1/(\s{\tilde{k}}-m_c)$ in Eq.~(\ref{eq:qed}) in the gluon field 
strength, which yields a series in differential
operators in the Fourier transform (`momentum') space. These act on
the momentum space `unity' in Eq.~(\ref{eq:qed}), which implies that the
right-most derivative vanishes.

To derive this representation, we start with the definition
\begin{equation}
\aver{\bar c^a_\al c^b_\be (x)} =
\matel{x}{\Big(\frac{1}{i\s{D}\!-\!m_c}\Big)_{ba}^{\be\al}}{x}
\label{eq:first}
\end{equation}
where $i\!\!\not\!\!\!{D}\!-\!m_c$ is the Dirac operator 
in the gluon background,
and we have shown the spinor indices $\al$ and $\be$ explicitly.
We then write the coordinate matrix element $\matel{x}{...}{x}$ of the
Dirac propagator in the momentum space: using \,$\langle k|x \rangle =
e^{ikx}$\, we have (the Lorentz indices are not shown for simplicity)
\begin{eqnarray}
\matel{x}{\frac{1}{i\s{D}\!-\!m_c}}{x} =
\matel{x}{\frac{1}{i\s{\partial}\!+\!\s{A}(x)\!-\!m_c}}{x} = 
\sum_{p,k} \langle x|p\rangle
\matel{p}{\frac{1}{i\s{\partial}\!+\!\s{A}(x)\!-\!m_c}}{k}\langle{k}|x
\rangle &=&\nonumber \\[0.2cm] 
\sum_{p,k} e^{i(p\!-\!k)x} 
\matel{p}{\frac{1}{\s{k}\!-\!m_c\!+\!\s{A}(i\partial_k)}}{k}\,.
\msp{30}&&
\end{eqnarray}
Since $\langle p|k\rangle=\delta(p\!-\!k)$, the latter
expression upon summation over $p$ recovers the definition of the
operator acting on the momentum-space unity in Eq.~(\ref{eq:qed}):
\beq
\sum_{p,k}
\matel{p}{f(k;\partial_k)}{k} = \sum_{p,k} f(k;\partial_k)\delta(p\!-\!k)=
\sum_{k}f(k;\partial_k) \cdot 1\;.
\label{unity}
\eeq

The second step is to apply the  Fock-Schwinger gauge 
$(x\!-\!x_0)\!\cdot\!A(x)=0$, $A(x_0)=0$. We shall
set the fixed point $x_0$ to zero and then (for details, see Ref.~(\cite{NSVZ})
\begin{equation} 
A_\mu(x) = \int_0^1 d\lambda\, \lambda x^\nu\, G_{\nu\mu}(\lambda x) \; .
\end{equation}
We can Taylor expand the field strength in powers of $\lambda x$
and integrate over $\lambda$. Due to the specific choice of the 
gauge condition the ordinary derivatives in the Taylor expansion 
can be replaced by the covariant derivatives, and the
expansion of the gauge potential takes the following form:
\begin{equation}
A_\mu(x) = \frac{1}{2}x^\al G_{\al\mu}-
\frac{1}{3}x^\al x^\lambda D_\lambda G_{\al\mu} + \ldots \; . 
\label{a18}
\end{equation} 
Using this explicit form for the gauge potential in the covariant
derivative in the Dirac operator and replacing $x_\al$ by 
$-i \partial_{k_\al}$ we arrive at the representation 
Eqs.~(\ref{eq:qed})-(\ref{a11}).

\section{Some elements of the 2-loop calculation $\alpha_s(m_c)/m_c$}

Here we outline the principal steps of the calculation 
that lead to the result in Eq. (\ref{90}).
The graphs are depicted in Fig \ref{fig:2}.

\begin{figure}[tb]
$$\epsfxsize=1.00\textwidth\epsffile{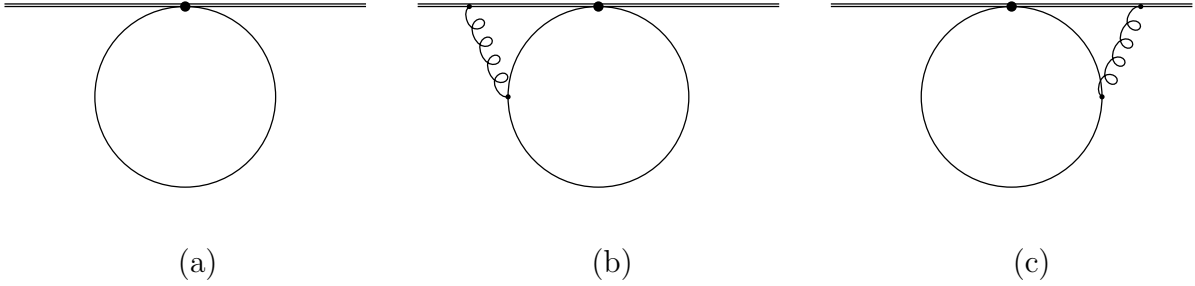}$$
\put(25,6){(a)}
\put(80,6){(b)}
\put(136,6){(c)}
\caption[]{The one-loop graph (a) and the relevant two-loop graphs
(b),(c); the external gauge field lines are not shown.
The two-loop graph (b) is depicted in detail in Fig.~\ref{fig:2}.}\label{fig:1}
\end{figure}

Our aim is to calculate the coefficients $C_{jk}(\mu)$ in Eq. (\ref{48}).
Using the universality of the OPE we can obtain those coefficients for
on-shell $b$-quarks with $p=(m_b,0,0,0)$ in the fixed gluon background,
where the matrix elements are simple functions of the gluon fields.

There are six possible ways to attach a hard gluon to the one-loop
graph in Fig \ref{fig:1}a; only two contribute at order $1/m_c$. The
gluon has to connect the $b$-quark line with the $c$ quark in the
loop,  Fig.~\ref{fig:1}b,c, otherwise the symmetry of the graph is the
same as in the one-loop case and would bar a $1/m_c$ contributions. 

The sum of the two-loop graphs in Figs.~\ref{fig:1}b,c reads 
\begin{eqnarray}
&&\msp{-8.7} \matel{b(p)}{J_{12}}{b(p)} = \\
&&  \msp{.7} - g^2\! \int_{kq} {\rm Tr} 
[\Gamma_2 S_c(p\!+\!k)t^a\gamma^\mu S_c(q)]
\, \Delta_{\mu\nu}^{ab}(k)\, \bar{b}(p) \left(t^b\gamma^\nu 
S_b(k+p)\Gamma_1 + \Gamma_1 S_b(p\!-\!k)t^b\gamma^\nu \right)b(p) 
\nonumber
\end{eqnarray} 
with $J_{12}$ defined in Eq.~(\ref{49}) and the trace taken over the
$c$-quark spinor and color indices; the $s$-quark and gluon
propagators in the external field are given below,
Eqs.~(\ref{prop2})-(\ref{prop6}). 
We use the 
convention $\int_{y}=\int d^dy$ and $\int_k =  \int \frac{d^d k}{(2\pi)^d}$. 
In the heavy quark limit the $b$-quark propagator $S_b$ reduces to
\begin{equation}
S_b(p\pm k) = \frac{\s{p} \pm \s{k}+m_b}{(p\pm k)^2-m_b^2+i0}
\stackrel{m_b \!\to \infty}{\longrightarrow } 
\frac{1}{\pm k_0+i0} = \pm \,P\frac{1}{k_0}- i\pi\delta(k_0) \; .
\end{equation}
\begin{figure}[tb]
$$
\begin{array}{@{}l@{}}
\epsfxsize=1.00\textwidth\epsffile{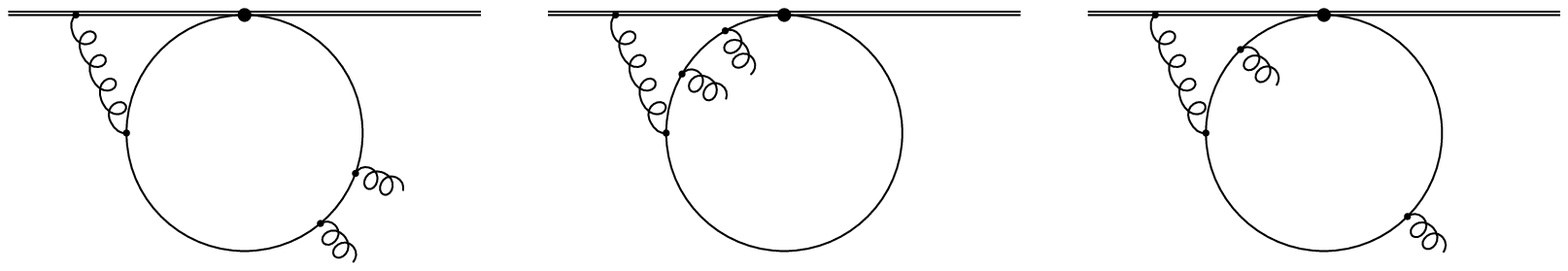} 
\end{array}
$$
\put(25,6){(a)}
\put(80,6){(b)}
\put(136,6){(c)}
$$
\begin{array}{@{}l@{}}
\epsfxsize=1.00\textwidth\epsffile{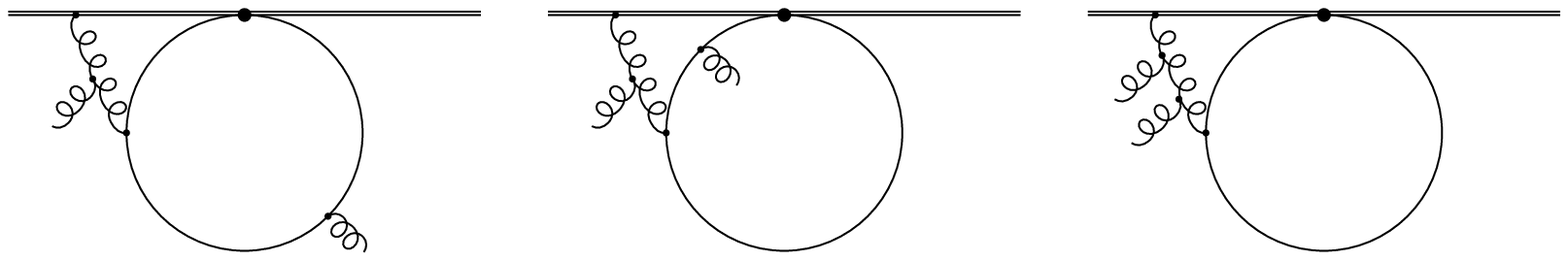}
\end{array}
$$
\put(25,6){(d)}
\put(80,6){(e)}
\put(136,6){(f)}
\caption[]{The graphs corresponding to Fig.~\ref{fig:1}b which contribute 
to order $\alpha_s(m_c)/m_c$ in the fixed-point gauge. The double
line shows the static $b$ quark and the loop stands for charm quark. The
two external gluon lines generically
represent the operators in Eqs.~(\ref{80}), as it comes out naturally
in this gauge.}\label{fig:2} 
\end{figure}
Only the $\nu\!=\!0$ vertex survives the static limit, and the
expression 
becomes
\begin{displaymath}
\matel{b(p)}{J_{12}}{b(p)} = \int_{kq} {\rm Tr}[\Gamma_2 
S_c(k+q)t^a\gamma^\mu S_c(q)]\, \Delta^{ab}_{\mu 0}(k) 
\; \bar{b} \left( i\pi\delta(k_0) \{t^b\!,\Gamma_1\} - 
P\mbox{{\large$\frac{1}{k_0}$}}\,[t^b\!,\Gamma_1]  \right ) b \nonumber
\end{displaymath}
It can be shown that the contribution of the principal value piece 
$P\frac{1}{k_0}$ vanishes upon integration; we have explicitly verified 
this in our calculation.

As the next step we need to put the appropriate propagators in the 
external field and to project out the operators in Eq.~(\ref{80}). 
As discussed in the previous Appendix, **for external gluon fields we use the background field 
method in the Fock-Schwinger gauge $x \!\cdot\! A(x)\!=\!0$ where 
the gauge potential is expanded directly in the gauge invariant 
field strength;
c.f. Ref \cite{NSVZ} for a pedagogical introduction. 
For internal gluon fields a different gauge may
be used; we adopt the Feynman gauge for it.

The gluonic operator parts in Eq.~(\ref{80}) carry dimension four and
may either come from a single propagator, or from two separate
propagators each contributing one gluonic field strength of mass
dimension two. It is most economic to choose the fixed
point to be the position of the local four-quark operator. 
With this choice the $b$-quark propagator does not interact with
soft gluons because the Wilson line is unity in this gauge; other
soft contributions to the $b$ quark propagator vanish in the heavy quark 
limit. There remain six graphs shown in Fig.~\ref{fig:2} 
out of ten possible combinatorial possibilities for Fig.~\ref{fig:1}b,
plus similar diagrams corresponding to Fig.~\ref{fig:1}c. 
They will be examined below.

\subsubsection*{Propagators in the external field}

The massless fermion propagator in the external field was given in
Ref.~\cite{NSVZ}, Eqs.~(2.30)-(2.31);\footnote{The $DDG$ term 
in $S^{(4)}$ was omitted in \cite{NSVZ}, Eq. (2.31), since the authors aimed
at calculating only the gluon condensate contribution $\aver{G^2}$.}  
the gauge potential propagator in the external field was also 
presented in \cite{NSVZ} for the case where one of the 
endpoints is the fixed point, Eq.~(2.35). 
For the fermion propagator we need to include mass.  
The diagrams in Fig.~\ref{fig:2} also include gluon Green function between 
two general points. Both generalizations are not difficult. 

We arrange the propagators in increasing dimension of the external 
gluon fields:
\begin{eqnarray}
\nonumber
\Delta_{\mu\nu}^{ab}(x,y) \msp{-3}&=&\msp{-3} -i\,\aver{T A_\mu^a(x)A_\nu^b(y)}  
= \Delta^{(0)ab}_{\mu\nu}(x,y) + 
\Delta^{(2)ab}_{\mu\nu}(x,y) + \Delta^{(3)ab}_{\mu\nu}(x,y)  + 
\Delta^{(4)ab}_{\mu\nu}(x,y) \\[3mm]
S(x,y) \msp{-3}&=&\msp{-3} 
-i\,\aver{T\psi(x)\bar{\psi}(y)}_A  = S^{(0)}(x,y) + 
S^{(2)}(x,y) + S^{(3)}(x,y) + S^{(4)}(x,y)\;,
\label{prop2}
\end{eqnarray}
The propagators are obtained using the background field Lagrangian.
With the fixed point $x_0\!=\!0$ the used terms for the 
fermion propagator are 
\begin{eqnarray}
S^{(0)}(x,y)\msp{-3} &\equiv&\msp{-3} S(x\!-\!y) = 
\int_p e^{-i p\!\cdot\! (x\!-\!y)} S(p) \equiv
\int_p \, e^{-i p\!\cdot\! (x\!-\!y)} \frac{\s{p}+m}{p^2\!-\!m^2}\;, \\[0.2cm]
S^{(2)}(x,y) \msp{-3}&=&\msp{-3} -\frac{1}{2} G_{\al\be}(0) 
\left\{ \begin{array}{l} (i\partial_{p}+x)_\al \\ (-i\partial_{q}+y)_\al \end{array} \right\}
\int_p e^{-i p \cdot(x\!-\!y)}S(p)\ga_\be S(q) \nonumber \;, \\[0.2cm]
S^{(4)}(x,y) \msp{-3}&=&\msp{-3} \frac{1}{4} G_{\al\be}G_{\gamma\delta}  
\left\{ \begin{array}{l} (-i\partial_{p}\!-\!x)_\al(-i\partial_{p+q}\!-\!x)_\gamma \nonumber \\ 
(-i\partial_{q+k}+y)_\al(-i\partial_{k}+y)_\gamma  \end{array} \right\}
\int_p e^{-ip \cdot (x\!-\!y)} S(p)\gamma_\be S(q) \gamma_\delta S(k) \nonumber  \\
\msp{-3}&-&\msp{-3} \frac{1}{8} \nabla_\al \nabla_\be G_{\gamma\delta}
\left\{ \begin{array}{l} (i\partial_{p}+x)_\al(i\partial_{p}+x)_\be(i\partial_{p}+x)_\gamma \nonumber  \\ 
(y\!-\!i\partial_{q})_\al(y\!-\!i\partial_{q})_\be(y\!-\!i\partial_{q})_\gamma \end{array} \right\} 
\int_p e^{-ip \cdot (x\!-\!y)} S(p)\gamma_\delta S(q)\;,  \nonumber
\label{prop4}
\end{eqnarray}
and for gluons 
\begin{eqnarray}
\Delta_{\mu\nu}^{\!(0)\,ab}(x,y) \msp{-3}&=&\msp{-3} 
g_{\mu\nu\,}\delta^{ab}\int_p 
e^{-i p \!\cdot\!(x\!-\!y)}\frac{-1}{p^2}\;, \\[0.2cm]
\Delta_{\mu\nu}^{(2)ab}(x,y) \msp{-3}&=& \msp{-3}
-2 G_{\mu\nu}^{ab}(0) \int_p
\,e^{-i p \!\cdot\!(x\!-\!y)}\frac{1}{p^4}\;, \nonumber \\
                             \msp{-3}&+&\msp{-3} 
i g_{\mu\nu}G_{\al\be}^{ab}(-i\partial_k+y)_\al \int_p e^{-i p \cdot(x\!-\!y)}
\frac{p_\be}{p^2}\frac{1}{k^2}\;,  \nonumber \\[0.2cm]
\Delta_{\mu\nu}^{(4)ab}(x,y) \msp{-3}&=&\msp{-3} 
-4 G_{\mu\la}^{ac}G_{\la\nu}^{cb} \int_p e^{-i p \cdot(x\!-\!y)} \frac{1}{p^6} 
\nonumber \\
    \msp{-3}&-&\msp{-3}  
\nabla_\al \nabla_\be G_{\mu\nu}^{ab}(y\!-\!i\partial_k)_\al
(y\!-\!i\partial_k)_\be
\int_p e^{-i p \cdot(x\!-\!y)} \frac{1}{p^2}\frac{1}{k^2}  \nonumber \\
   \msp{-3} &-&\msp{-3} 
\frac{1}{4} g_{\mu\nu} G^{ac}_{\al\la}G^{cb}_{\la\delta}  
(y\!-\!i\partial_k)_\al(y\!-\!i\partial_k)_\delta 
\int_p e^{-i p \cdot(x\!-\!y)} \frac{1}{p^2}\frac{1}{k^2} \nonumber \\
   \msp{-3} &+&\msp{-3} 
\frac{i}{4} g_{\mu\nu} (\nabla_\al\nabla_\be G_{\gamma\delta})^{ab} 
(y\!-\!i\partial_k)_\al(y\!-\!i\partial_k)_\be(y\!-\!i\partial_k)_\gamma
\int_p e^{-i p \cdot(x\!-\!y)} \frac{p_\delta}{p^2}\frac{1}{k^2}\,.
\nonumber 
\label{prop6}
\end{eqnarray}
For the fermion propagators we gave 
different possible representations which can be chosen pursuant to
calculational convenience, see below. The momenta $q$ and $k$ must be
set equal to $p$ after taking derivatives. For the gluon fields the
matrix notation $G^{ac}_{\mu\nu} \!=\! f^{abc}G^b_{\mu\nu}$ is
assumed.  In fact, one can do without a coordinate representation if
one of the endpoints of the propagator coincides with the fixed
point. This is the case for the $c$ quark lines, but not for the gluon
propagator in Fig.~\ref{fig:2}. $S^{(3)}$ or $\Delta^{(3)}$,
respectively, do not contribute, and we do not give $S^{(4)}$ and
$\Delta^{(4)}$ for brevity.

After putting in the propagators and
evaluating the traces the needed integral takes the generic form 
\begin{equation}
\label{eq:intype}
I_{nm}(\alpha,\beta,\gamma)=\int_{kq} \delta(k_0) 
\frac{k_{\mu_1}\dots k_{\mu_n} q_{\nu_1}\dots 
q_{\nu_m}}{(k^2)^\alpha (q^2\!-\!m^2)^\beta
((k+q)^2\!-\!m^2)^\gamma}\:, \qquad 0 \!\leq \! m+n \!\leq \!6 \; .
\end{equation}
Those integrals are in fact easily calculated in arbitrary dimensions
$d$. Rewriting the propagators as  
$(q^2\!-\!m^2)^a = (-1)^a (\vec{q}^{\,2} \!+\! (m^2\!-\!q_0^2))^a$  
suggests integrating over
${\rm d}^{d\!-\!1}q$  and ${\rm d}^{d\!-\!1}k$\, first.
We combine the denominators by introducing a Feynman-Schwinger 
parameter which is 
integrated over at the end yielding the Euler Beta-function.
The variable $k_0$ is fixed by the $\delta$-function and the remaining
integration over $q_0$ is elementary. 

For the first step the following master integral is needed:
\bea
\nonumber
&& \msp{-10} J(a,b,c;d)=\int \frac{{\rm d}^d k}{(2\pi)^d} \,
\frac{{\rm d}^d q}{(2\pi)^d}\:
\frac{1}{(k^2)^a (q^2\!+\!m^2)^b ((k\!+\!q)^2\!+\!m^2)^c}\,=\\[0.2cm] 
&& \msp{5}\frac{1}{(4\pi)^d} \,
\frac{\Gamma(\mbox{$\frac{d}{2}$}\!-\!a)
\Gamma(a\!+\!b\!+\!c\!-\!d)}{\Gamma(b)\Gamma(c)
\Gamma(\mbox{{\small$\frac{d}{2}$}})}\,
B(a\!+\!b\!-\!\mbox{$\frac{d}{2}$},a\!+\!c\!-\!\mbox{$\frac{d}{2}$})\:
(m^2)^{d\!-\!(a\!+\!b\!+\!c)}
\label{masterint}
\eea
and, therefore the scalar integral in Eq.~(\ref{eq:intype})
is
\bea
\nonumber
&& \msp{-10} I_{00}(a,b,c;d\!+\!1)=\int \frac{{\rm d}^d k}{(2\pi)^d} \,
\frac{{\rm d}^d q}{(2\pi)^d} \,
\frac{{\rm d}q_{0}}{2\pi}\:
\frac{(-1)^{a+b+c}}{(k^2)^a (q^2\!+\!(m^2-q_0^2))^b ((k\!+\!q)^2+(m^2-q_0^2))^c}\,=\\[0.2cm] 
&& \msp{2} \frac{i(-1)^{a+b+c}}{(4\pi)^{d+\mbox{{\footnotesize$\frac{1}{2}$}}}}\, 
\frac{\Gamma(\mbox{$\frac{d}{2}$}\!-\!a)\Gamma(a\!+\!b\!+\!c\!-\!d\!-\!
\mbox{{\small$\frac{1}{2}$}})}{\Gamma(b)\Gamma(c)
\Gamma(\mbox{$\frac{d}{2}$})}\,
B(a\!+\!b\!-\!\mbox{$\frac{d}{2}$},a\!+\!c\!-\!\mbox{$\frac{d}{2}$})\:
(m^2)^{d\!-\!(a\!+\!b\!+\!c)\!+\!\mbox{{\footnotesize$\frac{1}{2}$}}} . 
\label{masterint2}
\eea
In view of proliferating combinatorial possibilities 
we implemented the evaluation of the tensor integrals algorithmically
starting from the scalar master integral Eq.~(\ref{masterint2}). 

The $c$-quark loop potentially has UV divergences and the gluon loop 
potentially has IR divergences.
Since the one-loop contribution to this order in $1/m_c$ vanishes, 
the divergences must disappear in the sum over all the graphs. 
The only
diagrams involving IR or UV divergent integrals are those from 
Figs.~\ref{fig:2}\,d,e,f which are those where a gluon is emitted from 
the hard gluon line. Explicit calculations yield that, in fact, the
divergences cancel in each individual graph.  

Furthermore, a Furry-type analysis of the symmetries of the graphs 
can be done. Since the Gell-Mann matrices have no definite 
transformation properties under transposition, one has to resort 
to Hermitian conjugation. This distinguishes the QCD analysis
from the QED case. The outcome is that the two graphs in 
Fig.~\ref{fig:2}d,e are equal; 
the two graphs in Fig.~\ref{fig:2}a,b are equal in the vector channel,
but not in the axial channel.
This is verified explicitly in the calculations.
Another nontrivial consistency check would be to introduce a 
generic Fock-Schwinger fixed point and to observe its disappearance 
from the final result. This would require, however more involved 
calculations and the evaluation of four more graphs.

\vspace*{5mm}

%%%%%%%%%%%%

\end{document}